\documentclass[usenatbib,usegraphicx]{mn2e}
\usepackage{graphicx}
\usepackage{placeins}
\usepackage{color}
\usepackage{mathrsfs}
\usepackage{pstricks}
  
\begin{document}
\title[The XMM LSS pipeline]{The XMM Large Scale Structure survey:\\
The X-ray pipeline and survey selection function}

\author[F.~Pacaud et al.]{F.~Pacaud$^1$\thanks{E-mail: pacaud@discovery.saclay.cea.fr}, 
M.~Pierre$^1$, A.~Refregier$^1$, A.~Gueguen$^1$, J.-L.~Starck$^1$, 
I.~Valtchanov$^2$\thanks{Present address: ESA, Villafranca del Castillo, Spain},
\newauthor A.M.~Read$^3$, B.~Altieri$^4$, L.~Chiappetti$^5$, P.~Gandhi$^6$, O.~Garcet$^7$, 
E.~Gosset$^7$,\newauthor  T.J.~Ponman$^8$, J.~Surdej$^7$\\
$^1$CEA/DSM/DAPNIA, Service d'astrophysique, F-91191 Gif-sur-Yvette, France\\
$^2$Astrophysics Group, Blackett Laboratory, 
Imperial College of Science Technology and Medicine, London SW7 2BW, UK\\
$^3$Department of Physics and Astronomy, University of Leicester, Leicester LE1 7RH, UK\\
$^4$ESA, Villafranca del Castillo, Spain\\
$^5$INAF IASF, Sezione di Milano `G.Occhialini', via Bassini 15, 20133 Milano, Italy\\
$^6$European Southern Observatory, Casilla 19001, Santiago, Chile\\
$^7$Institut d'Astrophysique et de G\'eophysique, Universit\'e de Li\`ege, All\'ee du 6 Ao\^ut, 
17, B5C, 4000 Sart Tilman, Belgium\\
$^8$School of Physics and Astronomy, University of Birmingham, Edgbaston, Birmingham, B15 2TT, UK}
\maketitle
\begin{abstract}
We present the X-ray pipeline developed for the purpose of the cluster
search in the XMM-LSS survey. It is based on a two-stage procedure via
a dedicated handling of the Poisson nature of the signal: (1) source
detection on multi-resolution wavelet filtered images; (2) source
analysis by means of a maximum likelihood fit to the photon
images. The source detection efficiency and characterisation are studied
through extensive Monte-Carlo simulations. This led us to define two
samples of extended sources: the C1 class that is uncontaminated, and
the less restrictive C2 class that allows for 50\% contamination. The
resulting predicted selection function is presented and the comparison
to the current XMM-LSS confirmed cluster sample shows very good
agreement. We arrive at average predicted source densities of about 7
C1 and 12 C2 per deg$^2$, which is higher than any available wide
field X-ray survey.  We finally notice a substantial deviation of the
predicted redshift distribution for our samples from the one obtained
using the usual assumption of a flux limited sample.
\end{abstract}
\begin{keywords}
surveys - X-ray: general - methods: data analysis - X-ray: galaxies: clusters 
- large-scale structure of Universe
\end{keywords}
\pubyear{2006}

\section{Introduction}

X-ray imaging is recognized to be one of the most sensitive and
reliable methods to detect galaxy clusters. The main reason for this
comes from the extended nature of the cluster emission, whose
intensity is closely related to the depth of the associated
potential well. Moreover, at high galactic latitude and
medium-deep X-ray sensitivity ($10^{-14}-10^{-15}~$erg~s$^{-1}$~cm$^{-2}$ in the
[0.5-2]~keV band), the mean source density\footnote{100 to 800 sources 
per deg$^2$ of which about one tenth is extended} is much lower than 
in the optical or NIR wavelengths.
Both aspects concur to significantly lower projection effects that
become critical in the optical bands above $z > 0.5$. The task of
discovering and characterizing X-ray clusters is, however, complicated 
by the Poisson nature of X-ray data combined with several instrumental
effects (PSF, vignetting, CCD patterns, X-ray and
particle background) that have to be disentangled from the intrinsic 
emission profile of the sources.

With its mosaic of overlapping $10^4$\,s XMM pointings, the XMM-LSS
survey has been designed to detect a significant fraction of the
cluster population out to z=1, over an area of several tens
of deg$^2$, so as to constitute a sample suitable for cosmological
studies \citep{xmmlss}. Trade-off in the survey design was depth
versus coverage, keeping within reasonable limits the total observing 
time.
The two major requirements of the X-ray processing were thus to
reach the sensitivity limit of the data in a statistically
tractable manner in terms of cluster detection efficiency, and to
subsequently provide the selection function of the detected
objects. 

\indent To achieve these goals, it was necessary to design a new two-step
X-ray pipeline, combining wavelet multi-resolution analysis and
maximum likelihood fits, both using Poisson statistics. This
substantial development was required, as our controlled tests on
simulated cluster fields revealed unsatisfactory performances for extended 
sources using the early versions of the official detection software 
provided by the XMM-SAS\footnote{XMM Science Analysis System,
http://xmm.vilspa.esa.es/sas/, for subsequent data analysis we used
v6.1 of this package} (see \citealt*{val2001}). 
Our approach follows the principles pioneered by \cite{160sqdeg} 
which were originally established for the ROSAT PSPC data, and 
that we have totally revised 
to optimally handle the complex XMM instrumental characteristics.

The present paper provides a detailed description of our pipeline
- a two year effort - , of its performances and of the resulting
computation of the selection function both for point-like and
extended sources. Section \ref{principle} describes the various steps 
and parameters of the procedure. Section \ref{simu} presents a global 
evaluation of the pipeline using Monte-Carlo image simulations. 
These are in turn used to define a system of classes for cluster 
candidate sources, allowing for various degrees of completeness or 
contamination. 
Finally, in section \ref{cosmo}, we present a case study for the 
computation of the survey selection function, relying on the pipeline 
source classification, in a standard cosmological context.

\section{Pipeline description}
\label{principle}
The pipeline proceeds in three steps:
\begin{enumerate}
\item Starting from raw observation data files (ODFs), calibrated 
event lists are created using the XMM-SAS tasks \texttt{emchain} and 
\texttt{epchain}. 
These are then filtered for solar soft proton flares and used to 
produce images.
\item The images are filtered in wavelet space, then scanned by a source 
detection algorithm set to a very low threshold to obtain a primary source 
list. 
\item Detailed properties of each detected source is assessed 
from the photon images using {\it \texttt{Xamin}}, a maximum 
likelihood profile fitting procedure. 
This package was designed for the purpose of the XMM-LSS survey, with 
the specific goal of monitoring in a clean and systematic way the 
characterization of extended X-ray sources and associated selection 
effects. 
\end{enumerate}

 \subsection{Image extraction}
 Once event lists have been created, proton flare periods are filtered
 following the method proposed by \citet{pratt2002}, i.e. using the
 light curves of high-energy events (10-12~keV for MOS, 12-14~keV for
 PN).  Histograms of each light curve, binned by 104 seconds, are
 produced and fitted by a Poisson law to determine the mean of the
 distribution, $\lambda$.  We then apply a 3$\sigma$ threshold, so
 that time intervals where the emission exceeds
 $\lambda+3*\sqrt{\lambda}$ are thrown out as contaminated.

 Images of 2.5\arcsec/pix containing single and double events are then 
 produced using \texttt{evselect} in each of the 5 energy bands: 
 $[0.3$-$0.5]$, $[0.5$-$2]$, $[2$-$4.5]$, $[4.5$-$10]$ and $[2$-$10]$~keV.

 \subsection{Source detection}   

 In order to maximize detection rates, and provide good input to
 the maximum likelihood fit for both point-like and extended sources
 within an acceptable computation time, we follow the prescription of 
 \citet{val2001}, extensively tested over numerical simulations, to 
 use a mixed approach combining wavelet filtering of the images and 
 detection with a procedure initially developed for optical images 
 (\texttt{SExtractor}).
 
 In each band, the 3 EPIC detector images are co-added and the resulting 
 image is filtered using the wavelet task \texttt{mr\_filter} from 
 the multi-resolution package \texttt{MR/1} \citep*{mr1}.
 This task incorporates a statistically rigorous treatment 
 of the Poisson noise which enables the removal of unsignificant signal 
 directly in the wavelet space using a thresholding algorithm. A subsequent 
 iterative image reconstruction process accurately recovers 
 the flux and shape of the relevant structures contained in the data. 
 The details of the procedure and an evaluation of its ability to properly 
 reconstruct faint sources in the Poisson regime are given in 
 \citet{starckpierre} and \citet{val2001}.

 The primary source catalogues are then derived running \texttt{SExtractor}
 \citep{sextra} on the filtered image. The use of this
 software is now possible because the multiresolution filtering has 
 removed most of the noise and produced a smoothed background. 
 To avoid border effects we restrict our analysis to the 
 inner 13\arcmin of the field\footnote{The centre of the pointing is
 computed as a sensitivity-weighted average of the optical axis
 positions of the three telescopes, taken from the exposure map header
 keywords XCEN and YCEN}.
 With our current settings, the software essentially proceeds in four steps.
 First the background level is iteratively estimated in image cells by 
 $3\sigma$ clipping and a full-resolution background map is 
 constructed by bicubic-spline interpolation. Sources are then identified 
 as groups of adjacent pixels matching an intensity level. The software 
 subsequently tries to split blended sources by re-thresholding at some 
 sublevels between the original threshold and the peak value of each source  
 and looking for features containing a significant amount of the flux in 
 the emission profile. Finally a detailed analysis of the source is performed: 
 isophotal analysis to determine source position and shape, and photometry 
 in a flexible elliptical aperture as defined in \citet{SEkron} and 
 \citet{SEinfante}. 
 
 Parameters of the source detection steps are summarized in Table \ref{param}. 
 
 \begin{table}
 \begin{center}
 \caption{Relevant parameters of the XMM-LSS pipeline detection stage.
 Note that the high \texttt{SExtractor} detection threshold does not 
 imply that we are being restrictive, but rather reflects the fact 
 that the software is run on already adaptively smoothed images.
 \label{param}}
 \begin{tabular}{lc}
 \hline
 Parameter & value\\
 \hline
 \multicolumn{2}{l} {\bf Event selection:} \\
 MOS event flag selection	&	\#XMMEA\_EM\\
 PN event flag selection	&	(FLAG \& 0x2fb002c)==0\\
 MOS patterns			&	[0:12]\\
 PN patterns			&	[0:4]\\
 \multicolumn{2}{l} {\bf Image:} \\
 Type 				& 	sky\\
 Configuration 			& 	co-addition of EPIC detectors\\
 Pixel size 			& 	2.5\arcsec\\
 \multicolumn{2}{l} {\bf MR/1:} \\
 Wavelet type 			& 	B-spline\\
 Transform algorithm 		& 	``\`a trou" \\
 Poisson noise threshold  	& 	$10^{-3}$ ($\sim$ Gaussian $3.1\sigma$)\\
 Lowest significant scale 	& 	$2$ pix.\\
 Highest significant scale \hspace{0.5cm}	& 	$256$ pix.\\
 \multicolumn{2}{l} {\texttt{SExtractor}:} \\
 Background cell side 		& 	64 pix.\\
 Background median filtering	&	4 cells\\
 Detection threshold		&	6$\sigma$\\
 Detection minimum area		&	12 pix.\\
 Deblending sub-thresholds	&	64\\
 Deblend min. contrast		&	0.003\\
 \hline
 \end{tabular}
\end{center}
\end{table}

 \subsection{Source validation and characterization: {\it \texttt{Xamin}}}
 
 At the end of the pipeline processing, all the sources detected 
 by \texttt{SExtractor} are analyzed by {\it \texttt{Xamin}} using the binned 
 photon images. 
 
 For each source, {\it \texttt{Xamin}} determines a model that 
 maximizes the probability of generating the observed spatial photon
 distribution. First, a point source model is tested, then an extended source 
 profile parametrized by a spherically symmetric $\beta$-model 
 \citep{cavaliere1976}:
 \begin{equation}
 	S_X(r) \propto \left[ 1 + \left(\frac{r}{r_c}\right)^2 \right]^{-3\beta + 1/2},
 \end{equation} 
 convolved with the XMM point spread function.
 As we generally don't have enough S/N with our data to estimate
 simultaneously both $r_c$ and $\beta$ (especially with a 2D fit, we
 decided to fix $\beta$ to the canonical value of $2/3$ that is widely
 used to model the X-ray emission profile of massive galaxy clusters.
 Similarly, fitting more sophisticated models (e.g. 
 elliptical) is not justified.
 The best fit parameters for both models are listed in output along with
 relevant parameters characterizing the source (see list of Table
 \ref{xaminpar}).

 \subsubsection{Likelihood model}
 \label{stat}
 The statistic used to assess the reliability of a given model is a 
 simplified version of the C-statistic \citep{cash}:
 \begin{equation}
 	C=2\sum_{i=1}^{N_{pix}} \left( m_i - y_i\ln{m_i}\right),
 \end{equation}
 where $y_i$ is the number of observed photons in pixel $i$, and $m_i$ is the 
 model value in that same pixel.
 In our specific case, the emission profile of a source is the product of its 
 normalization $N_{mod}=\sum_{i=1}^{N_{pix}} m_i$ and its spatial distribution 
 $d_i$, which are independent: $m_i=N_{mod}\times d_i$. The C-statistic thus 
 reads:
 \begin{equation}
 	C=2 \left( N_{mod} - N_{data}\ln{N_{mod}} \right) - 2\sum_{i=1}^{N_{pix}} \left(y_i\ln{d_i}\right),
 \end{equation}
 where $N_{data}=\sum_{i=1}^{N_{pix}}y_i$. Minimization of 
 the C-statistic with respect to $N_{mod}$ directly yields 
 $N_{mod}=N_{data}$ and we consequently decided to fix 
 $N_{mod}$ and use the statistic:
 \begin{equation}
 	E = - 2\sum_{i=1}^{N_{pix}} \left( y_i\ln{d_i}\right),
 \end{equation}
 which is equivalent to the C-statistic as far as parameter estimation is 
 concerned. 
 This formalism has the advantage of reducing the parameter space of the 
 fit by one dimension (the overall normalization). However, it should be 
 noted that the normalization term $2 \left( N_{mod} - N_{data}\ln{N_{mod}} \right)$
 that we have cancelled for the fit still impacts on the error
 budget, and has to be reintroduced while computing confidence ranges.
 
 Here, we stress that, despite the common terminology, 
 the $C$ (and $E$) statistics are not likelihood functions (which have the 
 dimension of a probability or probability density), but are actually 
 related to the true likelihood $\mathcal{L}$ by:
 \begin{equation}\label{likedef}
 	C = -2\times\log{\mathcal{L}} + B, 
 \end{equation}
 where $B$ is a constant. 
 
 As for the C-statistic, the increase of $E$ between its 
 best fit value ($E_{B.F.}$) and a model containing only background 
 (i.e. uniform distribution of the photons), which is often improperly 
 referred to as `detection likelihood', quantifies the significance 
 of a detection and is $\chi^2$ distributed in the limit of large number 
 of counts (see \citealt{cash}).
 From now on, we refer to this parameter as the {\it detection statistic}:
 \begin{equation}
 	DET\_STAT = 2 N_{data} \ln(N_{pix}) - E_{B.F.}.
 \end{equation} 
 Similarly, the significance of the estimated extension, can be assessed using
 an {\it extension statistic} (improperly referred to as `extension likelihood') 
 which compares the value of $E$ for the best fit point-like and extended source 
 models (once again $\chi^2$ distributed in the limit of large number of counts): 
 \begin{equation}
 	EXT\_STAT = {\left(E_{B.F.} \right)}_{point} - {\left(E_{B.F.} \right)}_{extended}.
 \end{equation}
 
 The interpretation of these statistics in terms of a detection/extension 
 probability using the $\chi^2$ limit depends on the number of fitted 
 parameters. 
 All the statistics are thus ultimately converted into 
 equivalent values that would correspond to a fit with two free 
 parameters yielding the same probability. This provides 
 a unique and well-defined link between our statistics and probability: 
 for any statistic S, $P = \exp^{-\frac{S}{2}}$.
  
 \subsubsection{Source processing}
 \label{xamin-process}
 For each source, a fitting box is extracted, the size of which
 depends on the \texttt{SExtractor} inputs (start with 3 times the
 estimated FWHM, with the added requirement to be 
 always at least 35\arcsec ).  
 The \texttt{SExtractor} pixel segmentation mask is used to flag out pixels 
 belonging to neighbouring sources included in the box. 
 This method works well both for source characterisation 
 and classification in our shallow exposures\footnote{we detect some 0.1 
 source per arcmin$^2$ for a PSF FWHM of 6"}, but one would ideally have to 
 implement a simultaneous fit of blended sources in very crowded fields (in
 development).
 
 The source models take into account all significant XMM instrumental
 effects: an image of the source emission profile is
 constructed\footnote{ For both point-source and extended source
 profiles, we use the MEDIUM PSF model from the XMM calibration data
 which is the only one that reproduces the strong distortions of the
 PSF at large off-axis angles} and normalized to the tested count
 rate\footnote{For extended sources, this count rate is actually
 required to match the integral of the profile to infinity, as a
 significant amount of the source flux can fall outside the fitting
 box}; this image is then multiplied by the exposure maps (taking into
 account vignetting, detection mask, quantum efficiency and the 
 azimuthal sensitivity variations due to the 
 anisotropic transmission from the Rating Grate Arrays)
 and a uniform background is added, whose level is set so as to match 
 our normalization requirement ($N_{mod}=N_{data}$). 
 Given the faint sources that we are analyzing, this 
 very simple background model is justified in absence of small scale 
 variations of the XMM background in the soft bands.  
 While the EPIC-PN detector is considered as an independent instrument, 
 both the EPIC-MOS detectors are assumed to provide the same count rate 
 for the source and are thus modelled as a single detector using the
 summed photon image and exposure map.
  
 Starting from the SExtractor outputs as a first guess, the statistic $E$ is 
 minimized using the simplex method AMOEBA \citep{NumRec}, for both the 
 point source and extended emission models.
 It takes some 10 minutes for {\it \texttt{Xamin}} to process the average 120 
 detections per pointing found by \texttt{SExtractor}. 
 The procedure output catalogue comprises 29 derived parameters in addition 
 to the 9 free parameters of the fits (4 for the point source model and 5 for 
 the extended profile). These are listed in Table \ref{xaminpar}. 
 
\begin{table}
\caption{{\it \texttt{Xamin}} output parameters. Notes: $^a$ listed in the catalogues 
for both point-like and extended profile fits, $^b$ issued for each of 
the three EPIC detectors. Free parameters of the fitting process are written in 
bold font.\label{xaminpar}}

\begin{tabular}{ll}
   \hline
   Parameter & content\\
   \hline
   CUTRAD 		& Size of the fitting box	\\   
   EXP$^b$     		& Mean exposure time in the box \\
   GAPFLAG$^b$	  	& Distance to nearest CCD gap	\\
   GAP\_NEIGHBOUR 	& Distance to nearest detected neighbour \\ & in the fitting box \\
   {\bf EXT}           	& Best fit core radius		\\
   EXT\_STAT      	& Extension statistic 		\\
   DET\_STAT$^a$       	& Detection statistic		\\
   {\bf{X\_IMA}},{\bf{Y\_IMA}}$^a$      & Best fit position in pixel	\\
   RA,DEC$^a$        	& Best fit sky coordinates	\\
   {\bf RATE\_MOS}$^a$ 		& EPIC-MOS count rate		\\
   {\bf RATE\_PN}$^a$   	& EPIC-PN count rate		\\
   SCTS\_MOS$^a$   	& Estimated source counts in MOS1+2	\\
   SCTS\_PN$^a$   	& Estimated source counts in PN	\\
   BG\_MAP\_MOS$^a$	& Background level in MOS1+2	\\
   BG\_MAP\_PN$^a$ 	& Background level in PN	\\
   PIX\_DEV$^a$ 		& Distance between input/output position\\
   N\_ITER$^a$   		& Number of AMOEBA iterations	\\
   \hline
\end{tabular}
\end{table} 
 
\section{Performance evaluation through Monte-Carlo simulations}
\label{simu}
\subsection{Description of the simulations}
To assess the quality of our data analysis, we performed extensive
Monte-Carlo simulations of $10^4$\,s XMM pointings with the software
\texttt{InstSimulation} \citep{val2001}. This procedure creates images
from a source list taking into account the main instrumental
characteristics (PSF, vignetting, detector masks, background, 
Poisson noise).
In the following, cluster searches (in simulations as well as real
pointings) are performed in the [0.5-2]~keV band, and stated count
rates or fluxes always refer to this band. Galaxy cluster 
emission is indeed barely detectable at higher energies in our low 
exposure pointings, because of the combined effect of the redshifted
bremsstrahlung exponential cut-off, the XMM drop in sensitivity and strong
particle background above 2~keV.

The PSF of the simulations is obtained from the XMM calibration files
MEDIUM model, while the azimuthally-averaged off-axis dependency at
1~keV is used to model the vignetting. When simulated, the particle
and photon background levels were taken from \citet{read2003}.  In
order to convert source fluxes to count rates, we assumed a constant
EPIC-PN to EPIC-MOS count rate ratio regardless of the source
spectrum. Note that in the following we will always refer to count
rates as the sum of MOS1, MOS2 and PN rates after vignetting
correction\footnote{in our $10^4$\,s pointings, $10^{-2}$~cts~s$^{-1}$
roughly corresponds on the optical axis to $100$~cts spread over the
three EPIC detectors and a flux of about
$9\times10^{-15}$~erg~s$^{-1}$~cm$^{-2}$ for both an AGN spectrum (a
power law SED with spectral index $\Gamma=1.7$) and a local 2~keV
cluster (thermal bremsstrahlung)}.  This means that for the same count
rate, a source is more easily detected near the centre than on the
border of the FOV.

Four kinds of simulations were performed:
\begin{table}
  \begin{center}
  \caption{List of cluster simulations. For each cluster core radius, the 
  number of simulated pointings performed for each count rate ($N_{point}$) 
  is given, as well as the number of simulated sources per pointing ($N_{Src}$)
  in the central 10\arcmin.\label{simtab}}
  \begin{tabular}{cccc}
  \hline
  Radius (\arcsec)  & Count rate & $N_{point}$ & $N_{Src}$.\\
  \hline
  \bf{10}  	  & 0.005 & 10 & 8  \\
  	   	  & 0.01  & 10 & 8  \\
  	   	  & 0.02  & 10 & 8  \\
  	   	  & 0.05  & 10 & 8  \\
  \vspace{0.2cm}  & 0.1   & 10 & 8  \\
  
  \bf{20}  	  & 0.005 & 10 & 8  \\
  	   	  & 0.01  & 10 & 8  \\
  	   	  & 0.02  & 10 & 8  \\
  	   	  & 0.05  & 10 & 8  \\
  \vspace{0.2cm}  & 0.1   & 10 & 8  \\

  \bf{50}  	  & 0.01  & 15 & 6  \\
  	   	  & 0.02  & 15 & 6  \\
  	   	  & 0.05  & 15 & 6  \\
  \vspace{0.2cm}  & 0.1   & 15 & 6  \\

  \bf{100} 	  & 0.02  & 30 & 4  \\
  	   	  & 0.05  & 30 & 4  \\
  	   	  & 0.1   & 30 & 4  \\
  \hline
  \end{tabular}
  \end{center}
\end{table}
\begin{itemize}
\item 30 pointings of $10^4$\,s containing only point sources. The flux 
distribution and source density is computed using the Log(N)-Log(S) from 
\citet{moret} down to $5\times10^{-16}$~erg~s$^{-1}$~cm$^{-2}$. The background values from 
\citet{read2003} were accordingly corrected for the contribution of 
point sources fainter than $4\times10^{-15}$~erg~s$^{-1}$~cm$^{-2}$ (approximative flux 
limit of their analysis). We assumed a random spatial distribution of the sources 
(therefore neglecting the known angular correlation among AGNs).  
\item 250 pointings of $10^4$\,s containing extended sources only 
($\beta$-model with fixed $\beta$=2/3) with simulated background. 
We simulated core radii of [10, 20, 50, 100] arcsec with count rates 
in the range [0.005, 0.01, 0.02, 0.05, 0.1] cts~s$^{-1}$ (see Table \ref{simtab} 
for the exact list). Spatial distribution of the sources was set 
at random so as to cover most of the area within 10\arcmin, with the 
extra requirement that sources do not overlap. 
\item 250 simulations with the same extended sources as previously, but 
injected into a real XMM-NEWTON $10^4$\,s pointing pertaining to the XMM-LSS 
(XMM Id: 0037980501), in order to estimate how extended source 
characterization is affected by the point source population.
\item 18 simulations containing close pairs of point-like sources
(separated by 20\arcsec) injected into a typical real XMM-NEWTON pointing to 
test the deblending capabilities of the pipeline. In the first 9 
simulations, 10 pairs of $3\times10^{-3}$~cts~s$^{-1}$ were added in the pointing,
while in the remaining ones, 5 pairs of $5\times10^{-2}$~cts~s$^{-1}$ were simulated.
These simulations are also relevant for cluster false detection rate as 
blended point sources may be characterized as `extended' sources. 
\end{itemize}

\begin{figure}
 \begin{center}
 \includegraphics[width=11.cm]{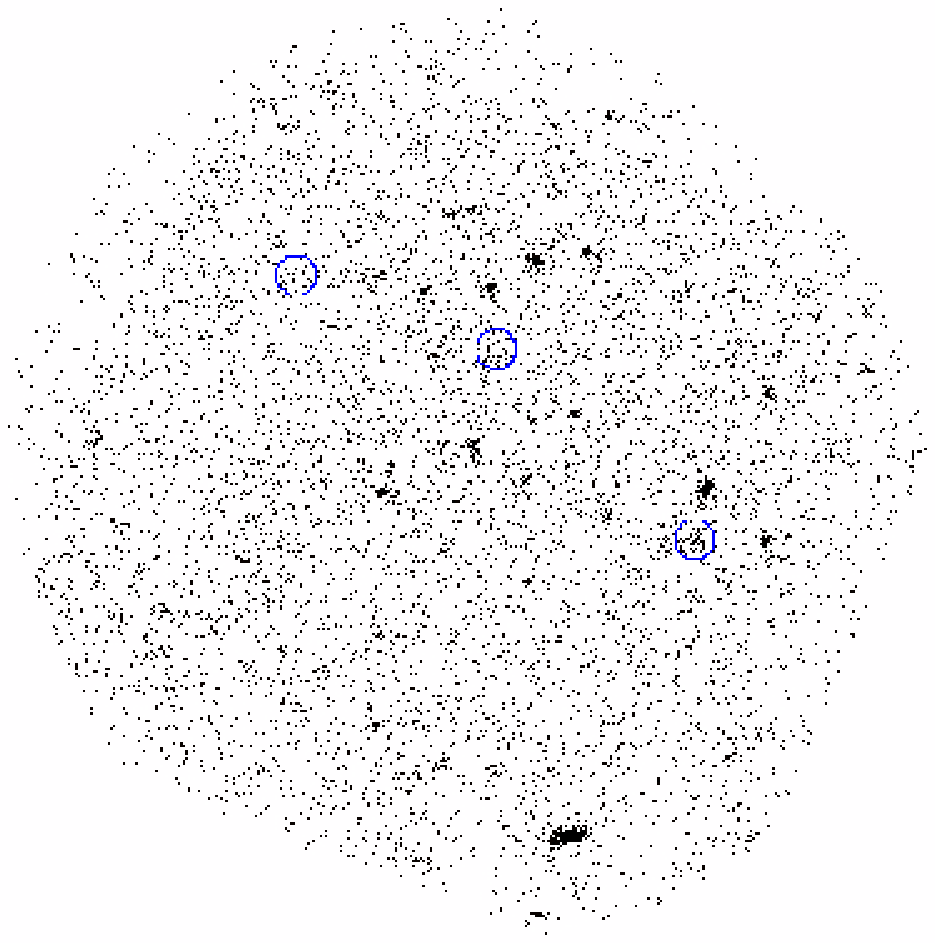}
 \includegraphics[width=11.cm]{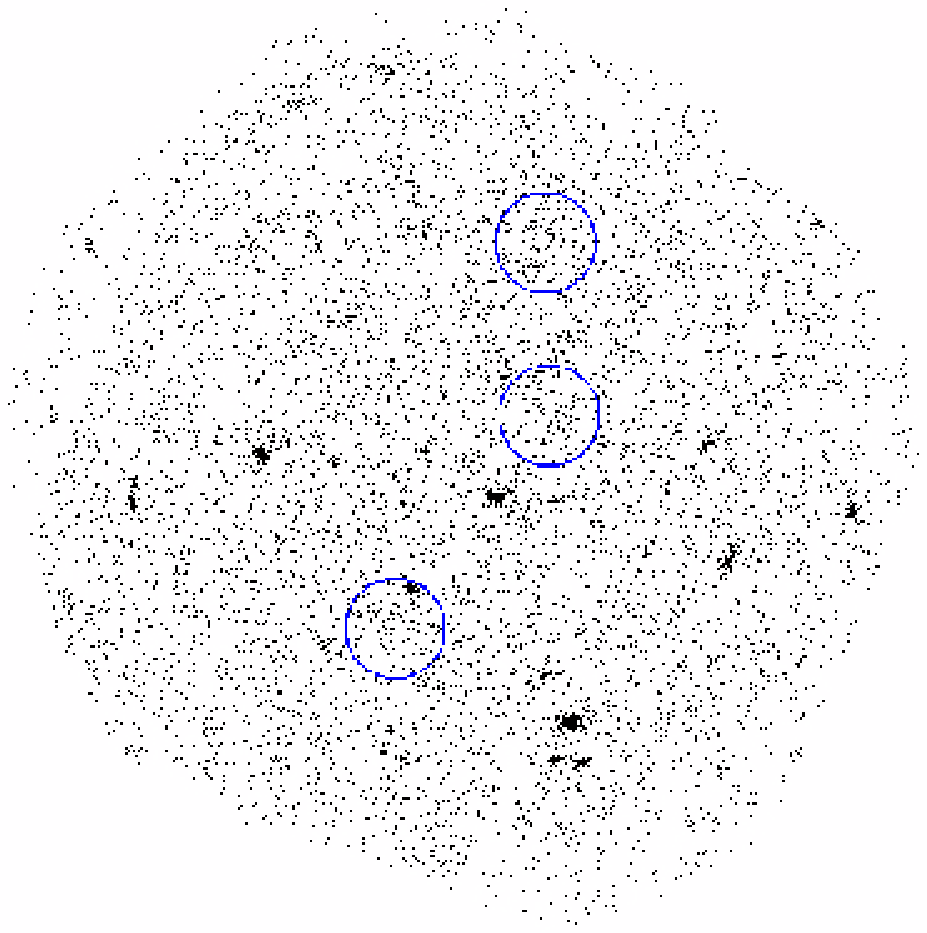}
 \caption{Examples of simulated $10^4$\,s XMM-NEWTON images (co-addition of the EPIC 
 cameras). Both contain point sources distributed following the X-ray 
 Log(N)-Log(S). Blue circles show the position of simulated clusters.
 Top: clusters have core radii of 20\arcsec and on-axis count rates of 
 (from top to bottom) 0.02, 0.01 and 0.03 cts~s$^{-1}$.
 Bottom: clusters have core radii of 50\arcsec and on-axis count rates of 
 (from top to bottom) 0.03, 0.02 and 0.05 cts~s$^{-1}$. Displayed clusters 
 are very faint (close to the detection limit) so as to illustrate the
 detection capabilities of the pipeline. 
 \label{simuimage} 
}
 \end{center}
\end{figure}

\begin{figure}
 \begin{center}
 \includegraphics[width=11.cm]{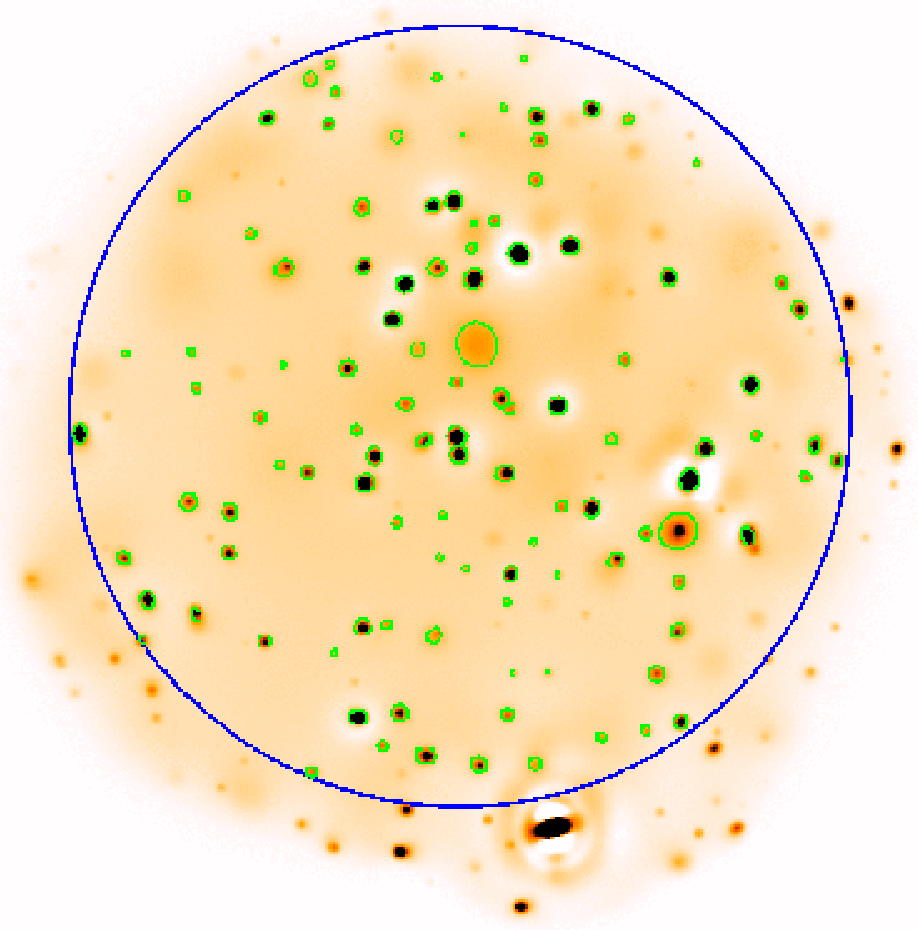}
 \includegraphics[width=11.cm]{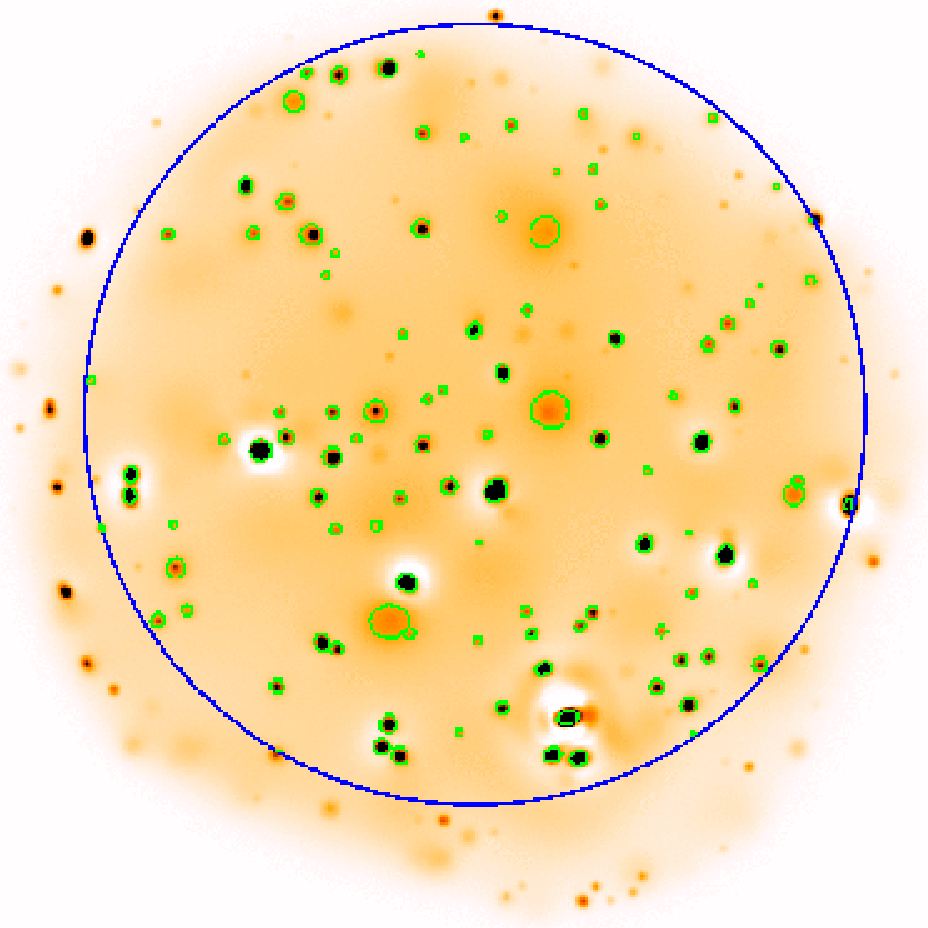}
 \caption{Wavelet images of fig. \ref{simuimage} simulations, overlayed with
 \texttt{SExtractor} catalogues. The blue circle shows the central 13\arcmin\ radius 
 of the FOV (centered on the mean optical axis) where \texttt{SExtractor} detections are 
 performed.
 \label{simuwave}}
 \end{center}
\end{figure}

\begin{figure}
 \begin{center}
 \includegraphics[width=11.cm]{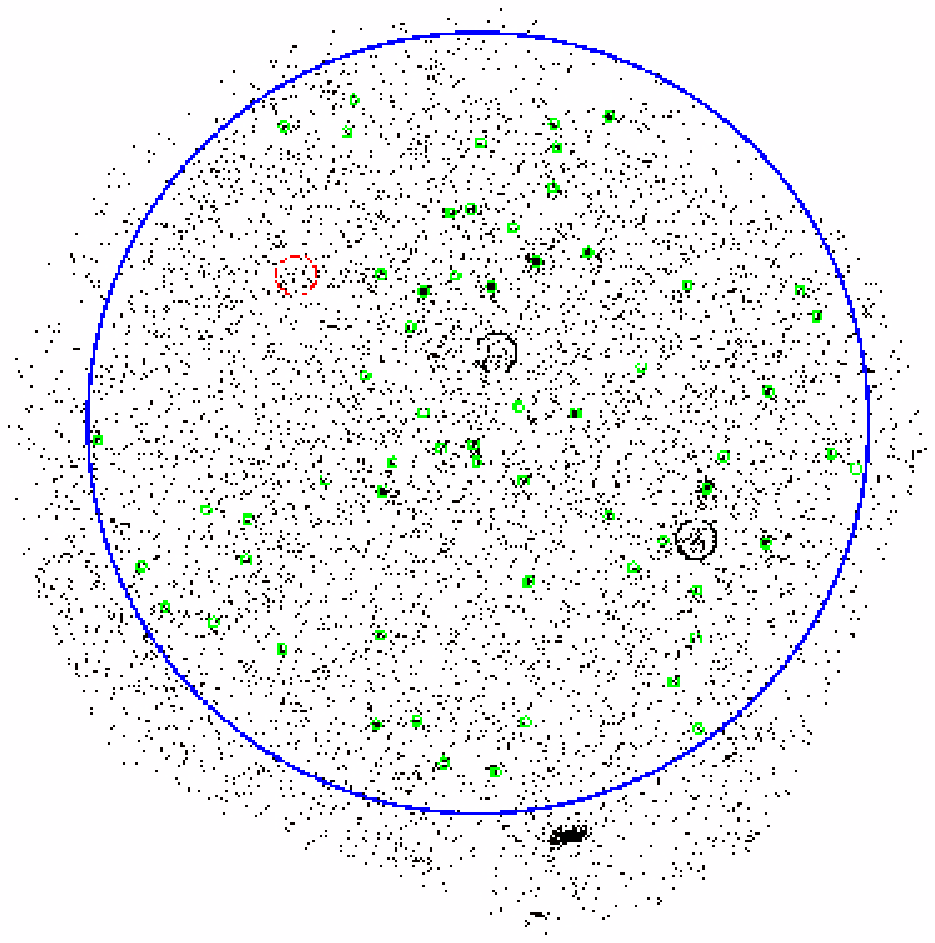}
 \includegraphics[width=11.cm]{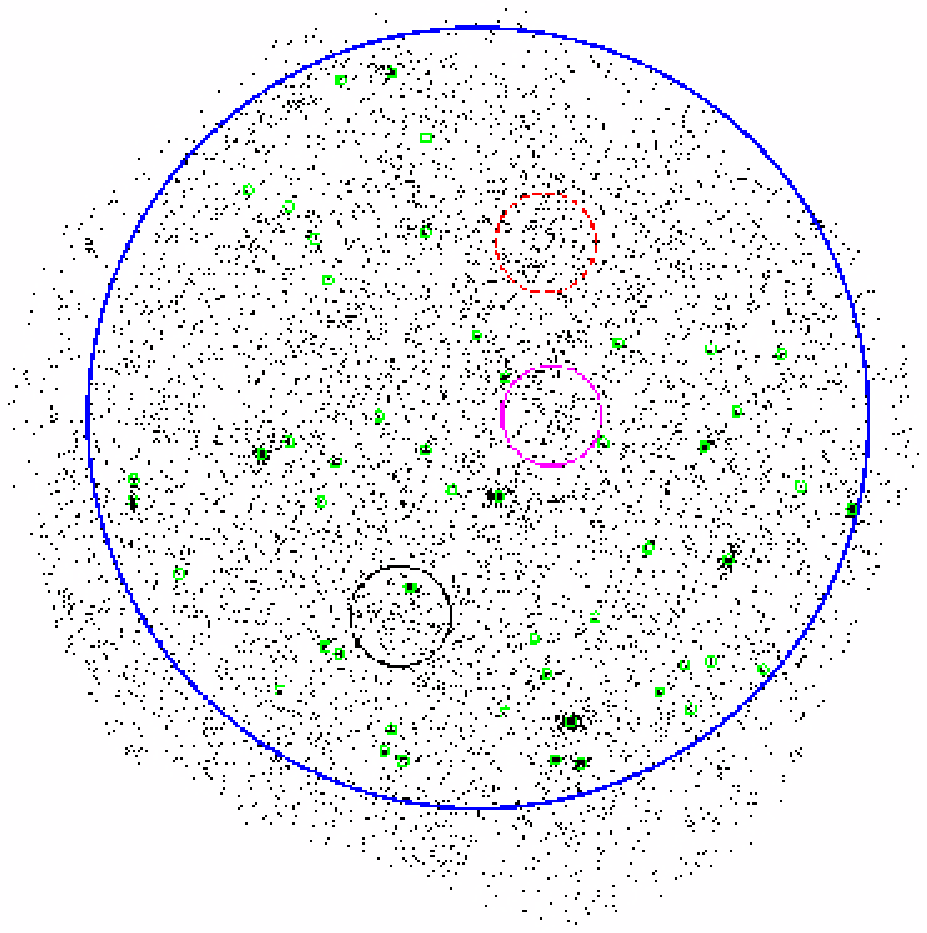}
 \caption{Raw images of fig. \ref{simuimage} simulations, overlayed with
 \texttt{\it Xamin} catalogues. The green 10\arcsec\ radius circles show the
 detected point sources (see fig. \ref{classtab} for the selection criteria); black
 and magenta circles show the C1 and C2 clusters respectively.
 Clusters not flagged by \texttt{\it Xamin} as C1 or C2 are indicated by red 
 dashed-line circles.
 \label{simuXamin}}
 \end{center}
\end{figure}

Examples of simulated images are given in figures \ref{simuimage}, 
\ref{simuwave} and \ref{simuXamin}.

All simulated images were analysed through steps (ii) and (iii) of our
pipeline (see \S\ref{principle}). Detected sources were then
cross-identified with the simulation inputs using a correlation radius
of 5~pixels for point sources and 15~pixels for extended ones. In the
following subsections, we will refer to {\it spurious} detections as
those that could not be cross-identified with any input source.

\subsection{Parameter estimation accuracy}
 \subsubsection{Extended sources}
\begin{figure}
\begin{center}
\vbox{\includegraphics[width=6.8cm]{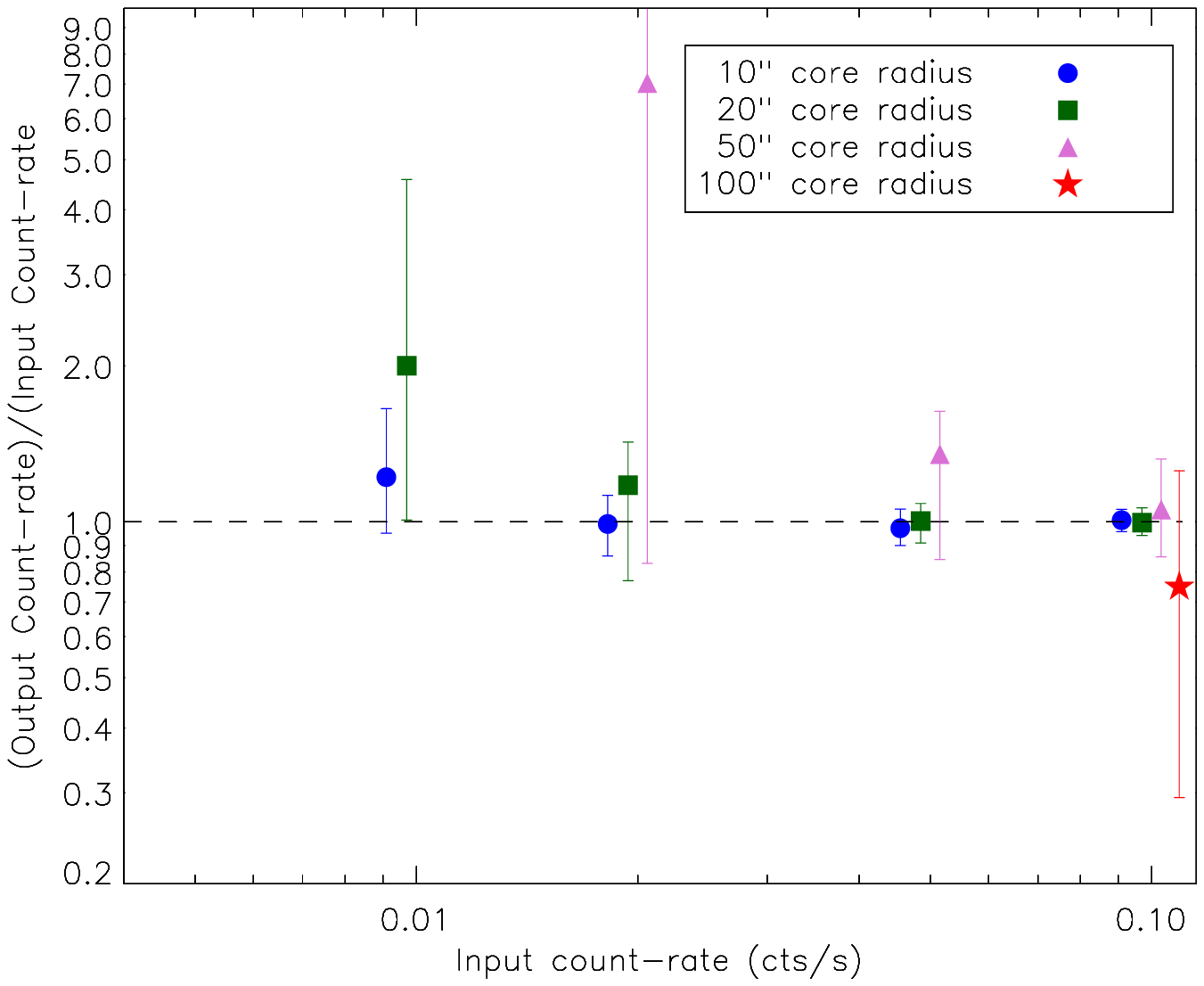}\hspace{.8cm}\includegraphics[width=6.8cm]{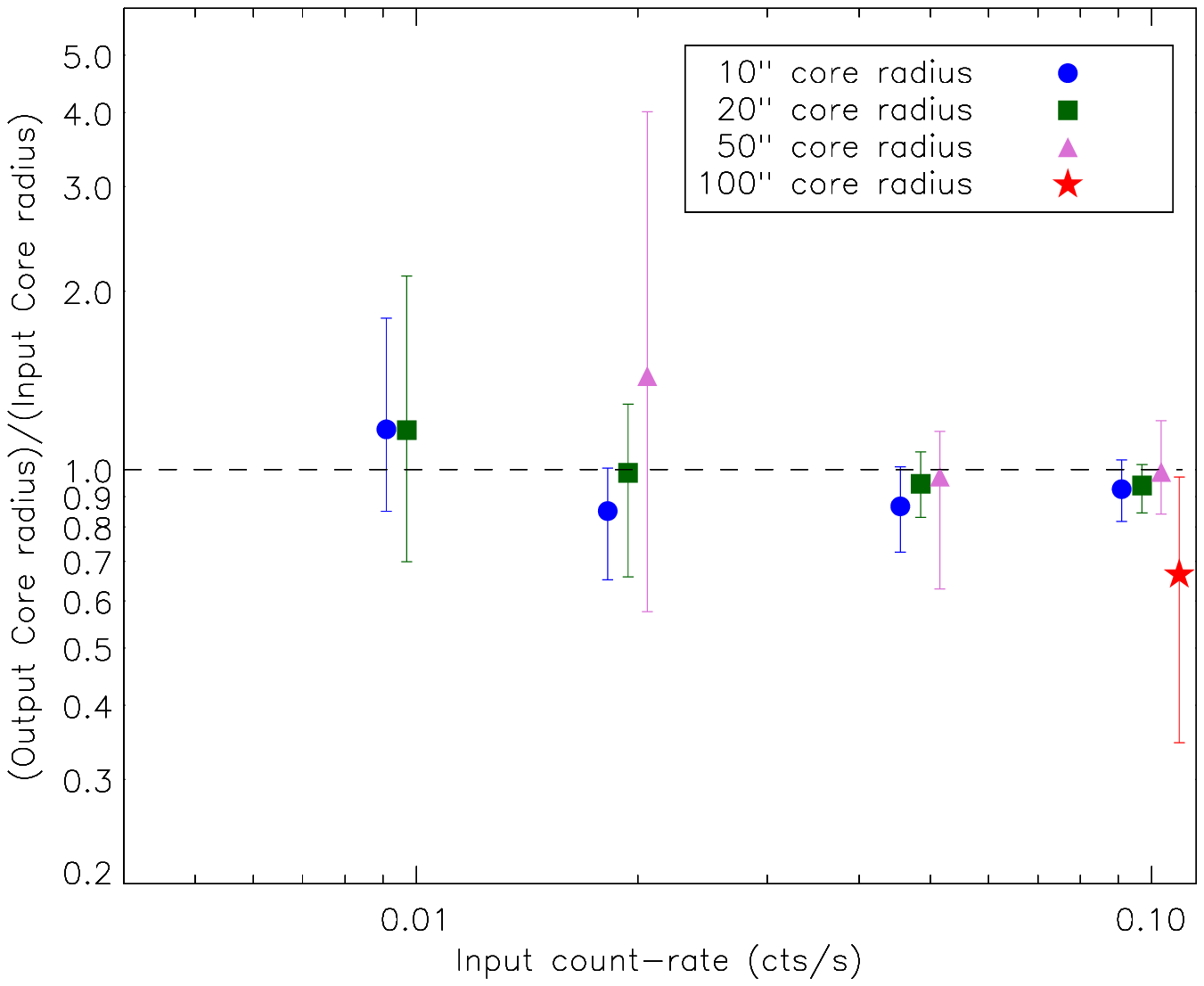}}
\vbox{\includegraphics[width=6.8cm]{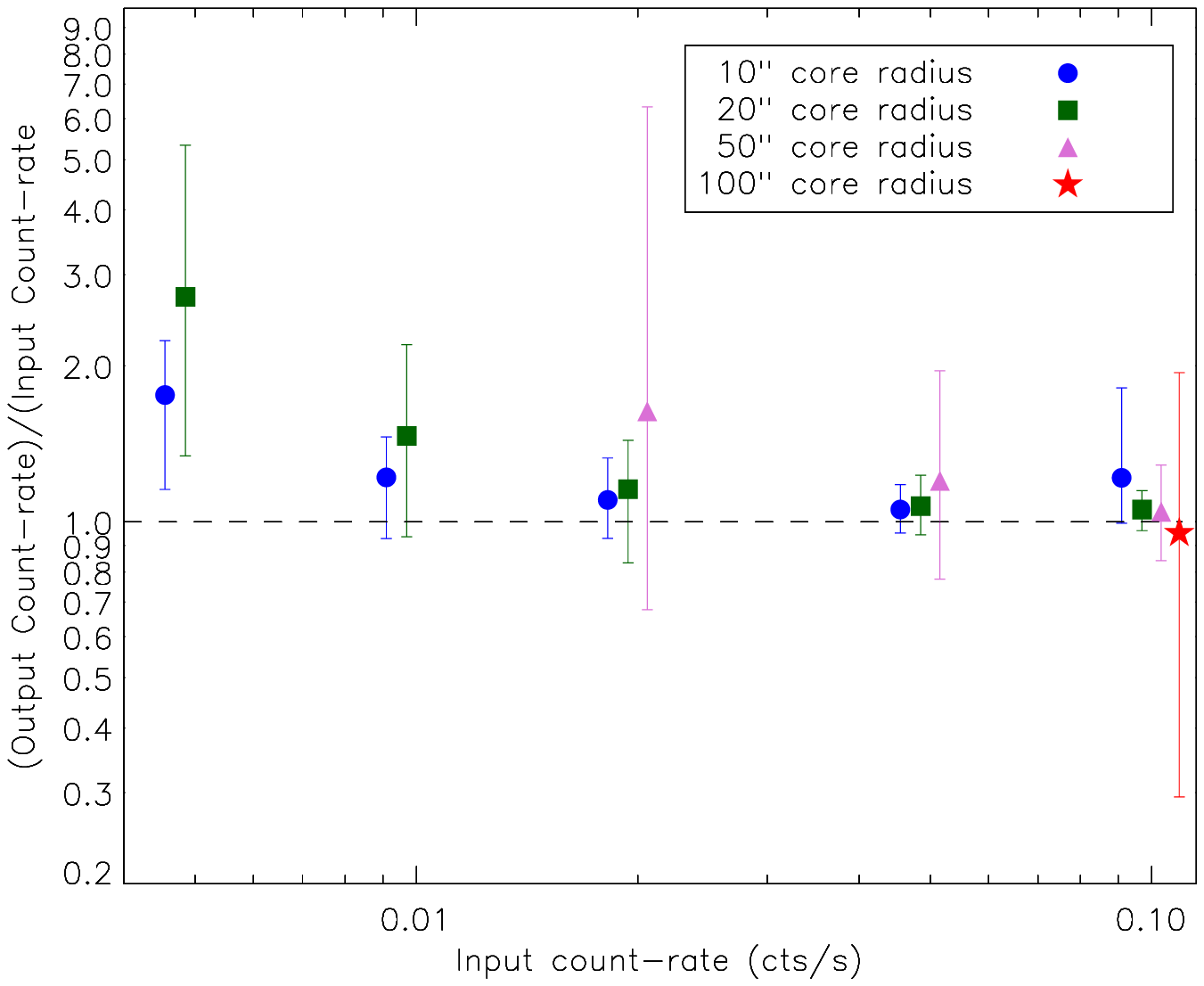}\hspace{.8cm}\includegraphics[width=6.8cm]{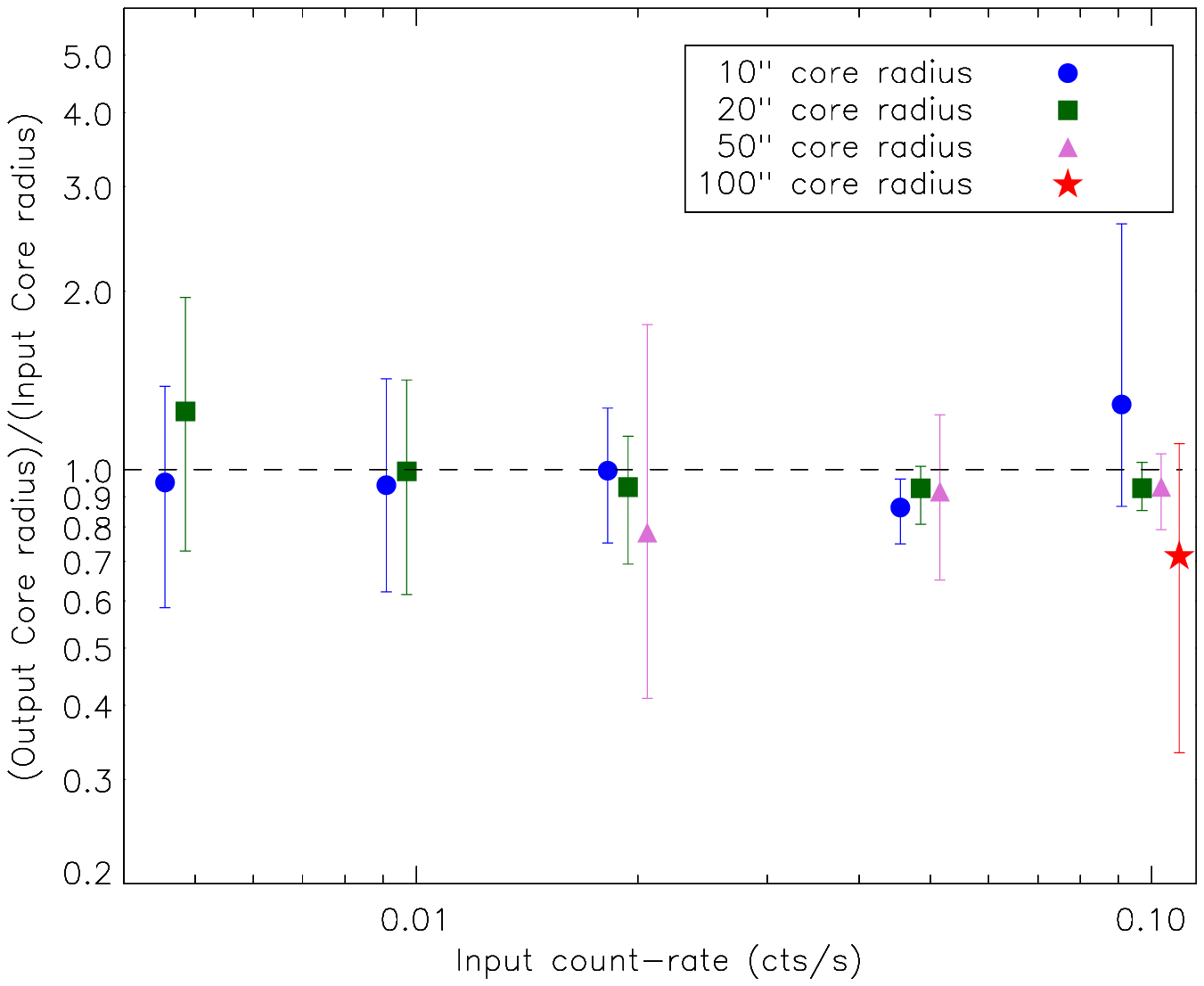}}
\caption{
Extended source characterisation with the XMM-LSS pipeline
within 10\arcmin\ of the FOV\label{extphoto} for $10^4$\,s exposures. For all plots, vertical bars 
show the standard deviation of measured points. Upper plots show the results for 
clusters in pointings without point sources, lower plots show the results when clusters 
are injected into real pointings. Left: photometry as a 
function of on-axis counts and core radius. Right: source 
extension measure as a function of on-axis counts and core radius.
In all the plots, only the bins encompassing at least 10 recovered sources 
are shown.}
\end{center}
\end{figure}

 Our simulations demonstrate that the mean photometry of extended 
 sources with {\it \texttt{Xamin}} is satisfactory in both pointings
 with or without point sources: 
 it is unbiased for bright sources with a mean dispersion 
 of about 20\% (see fig. \ref{extphoto}), while unavoidable the
 Eddington bias and scatter increase appear for fainter ones. 
 Count rates seem somewhat overestimated only for very faint or very extended 
 sources. The scatter increases slowly as count rates decrease but also with 
 increasing radius.
 
 Even when the clusters are injected into real pointings, the performances 
 remain correct up to 50\arcsec core radius. Knowing that, for a physical 
 core radius of $180h_{70}^{-1}$~kpc, the apparent core radii span the range
 $55\arcsec-22\arcsec$ for $0.2<z<1$ in $\Lambda$CDM, the goals of the 
 pipeline are fully met. 
 For very faint sources ($5\times10^{-3}$~cts~s$^{-1}$), the rates are 
 somewhat underestimated, which can be explained by the fact that
 only the central brightest part of the sources clearly emerges 
 from the background fluctuations.
 Above 50\arcsec core radii, the rates are somewhat overestimated.
 In addition, we notice surprisingly a weak increase of the detection 
 efficiency  of these sources when adding AGNs. 
 The simplest interpretation is that part of the very extended sources 
 found their emission contaminated by faint AGNs (that fall below our 
 detection or deblending capacity), and thus tend to 
 pass more easily the detection criteria of the pipeline, but with 
 erroneous photometry and core radius. 
 
 A second point to note comparing the left and right panels of figure
 \ref{extphoto} is that the photometry seems tightly correlated with
 extension measure accuracy. A poor modelling of the source emission
 profile logically yields incorrect count rate estimates, particularly
 for very extended sources where the flux is extrapolated far outside
 of the fitting box.  An inaccurate estimate of the source extensions
 is thus probably the reason for both the low count-rate and the high
 core radius photometry bias identified above.
     
 \subsubsection{Point sources}
 \label{pntconf}
\begin{figure}
\begin{center}
\begin{pspicture}(0,0)(16,6.5)
\vbox{
  \includegraphics[width=7.8cm]{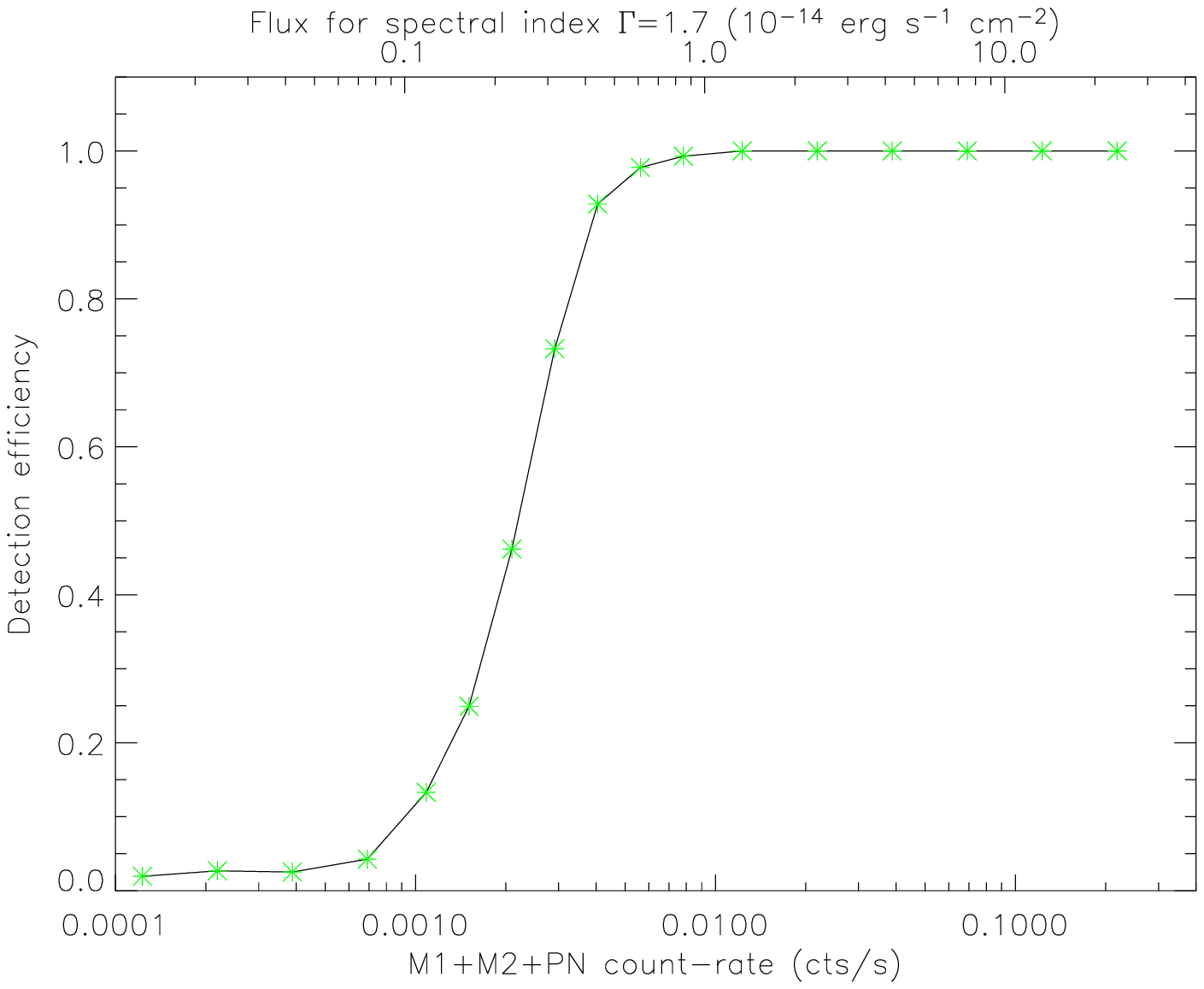}
  \includegraphics[width=7.8cm]{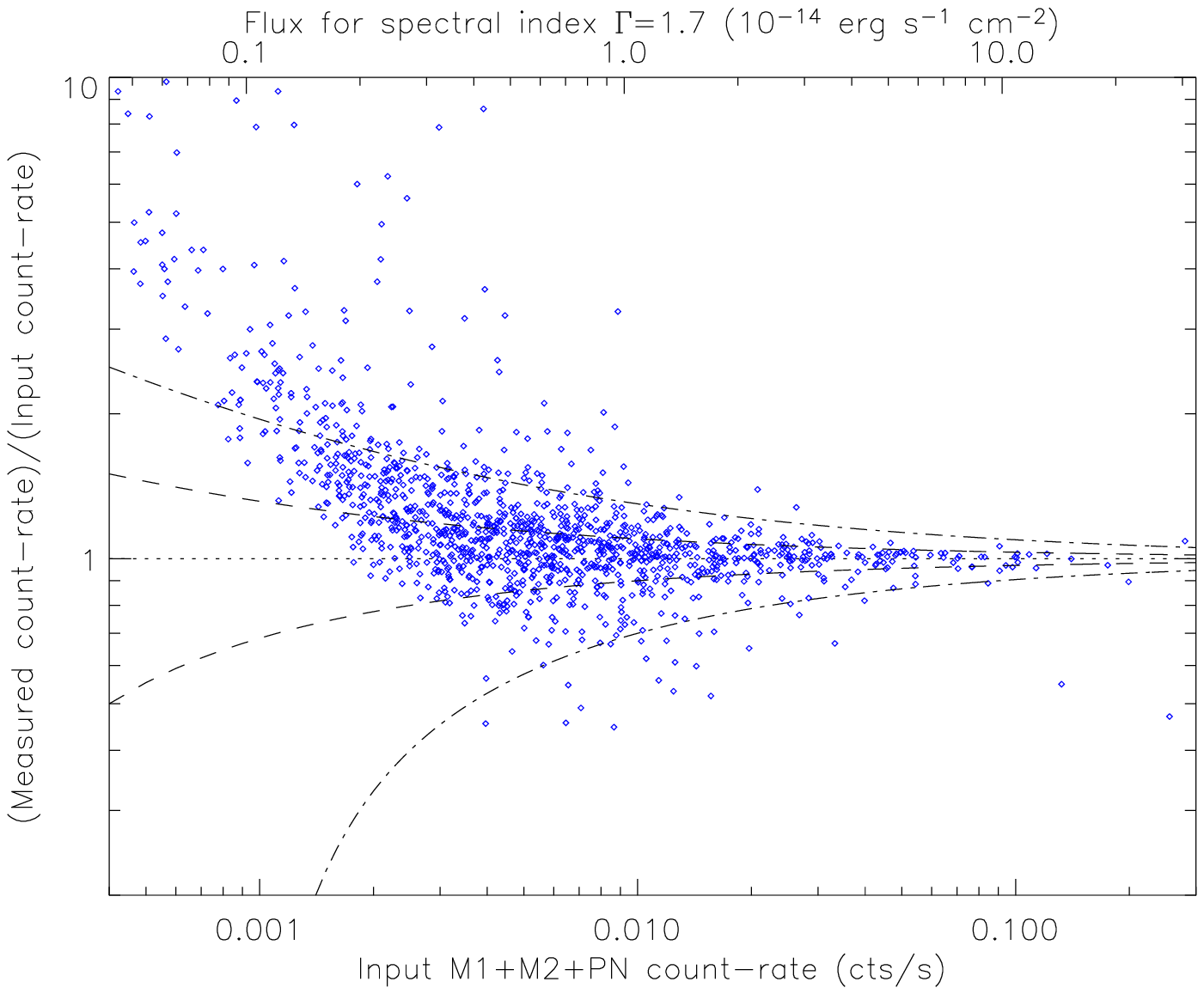}
  \rput(-8.9,1.2){\large (a)}
  \rput(-1,1.2){\large (b)}}
\end{pspicture}
\caption{
Point source analysis with the XMM-LSS pipeline 
within 10\arcmin\ of the FOV for $10^4$\,s exposures.\label{pntdet} Panel (a):
detection probability as a function of count rate. Panel (b): photometry, 
dashed and dotted-dashed lines show respectively the intrinsic $1\sigma$
and $3\sigma$ scatter expected from Poisson noise.
}
\end{center}
\end{figure}

 As shown in figure \ref{pntdet}b, the point source photometric
 dispersion is basically comparable to the spread due to Poisson noise down to 
 20 cts. At the faint tail of the distribution, a strong Eddington bias 
 appears.
 
 Another issue regards point source confusion. We used our set of
 close pair simulations to test the deblending efficiency. The results
 are quite satisfactory: all $5\times10^{-2}$~cts~s$^{-1}$ pairs are
 deblended, while more than 65\% of the $3\times10^{-3}$~cts~s$^{-1}$
 sources are also. This success rate cannot easily be reached by first
 step detection procedures based on sliding cells having a minimum
 size of 10\arcsec. This point is not only important for point source
 statistics, but also for cluster detection, in order not to consider
 blended point sources as a single extended one.

\subsection{Source classification using the simulations}
Source selection and estimation of the selection function in surveys
is always a complicated task and results from the necessary trade-off
between sample completeness and contamination.  For this purpose, we
explored the {\it \texttt{Xamin}} output parameter space by means of
our simulations in order to set well controlled extended/point-like
source selection criteria, and to estimate contamination by spurious
or misclassified sources.

\subsubsection{Point sources}

As AGNs represent more than 90\% of the extragalactic X-ray sources at
our sensitivity, we restrict ourselves to the estimation of the
spurious detection rate based on our point source simulations.  
As can be seen in figure \ref{class}a, a simple threshold of 15 in 
the detection statistic gives the best balance between contamination 
and completeness: 
at this threshold, some 40 to 50 real point-like sources are detected 
in each pointing within 10' of the FOV, for only 0.5 spurious ones.

The resulting detection efficiency as a function of count rate is
shown in figure \ref{pntdet}a.  The point-source flux limit (90\%
completeness) is about $4\times10^{-15}$~erg~s$^{-1}$~cm$^{-2}$ in
[0.5-2]~keV, but more than 50\% of the sources are detected down to
$\sim 2.5\times10^{-15}$~erg~s$^{-1}$~cm$^{-2}$.

\subsubsection{Extended sources}
\label{extclassdef}
\begin{figure}
\begin{center}
\vbox{\includegraphics[width=7.5cm]{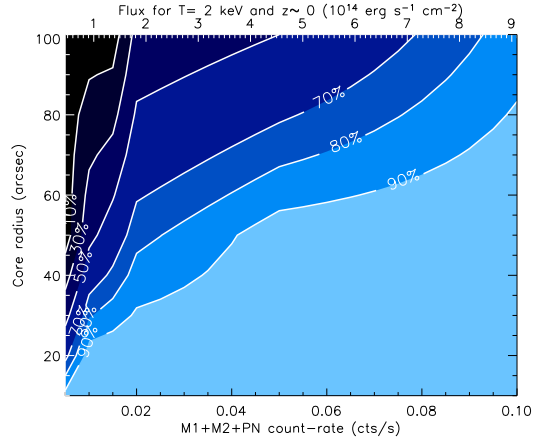}}
\caption{
Detection probability for extended sources by {\bf \texttt{SExtractor}};
this can be considered as the ultimate sensitivity with $10^4$\,s XMM images, 
but the contamination is maximal.\label{probSE}}
\end{center}
\end{figure}

Source selection is complicated for extended sources because these
objects are generally of lower surface brightness (see e.g. figure
\ref{simuimage}), and one does not only have to deal with spurious
detections, but also with contaminating misclassified point sources.
This task requires special care, keeping in mind the very cosmological
applications of the survey. 
Figure \ref{probSE} shows the fraction of extended 
sources that are detected by \texttt{SExtractor} in the primary catalogue 
as a function of flux and core radius. Our purpose is then to find a 
location in the {\em \texttt{Xamin}} output parameter space where the
majority of these sources are recovered while keeping the contamination 
rate to a manageable level.

\begin{figure}
\begin{center}
\begin{pspicture}(0,0)(15,7.)
\vbox{
  \includegraphics[width=7cm]{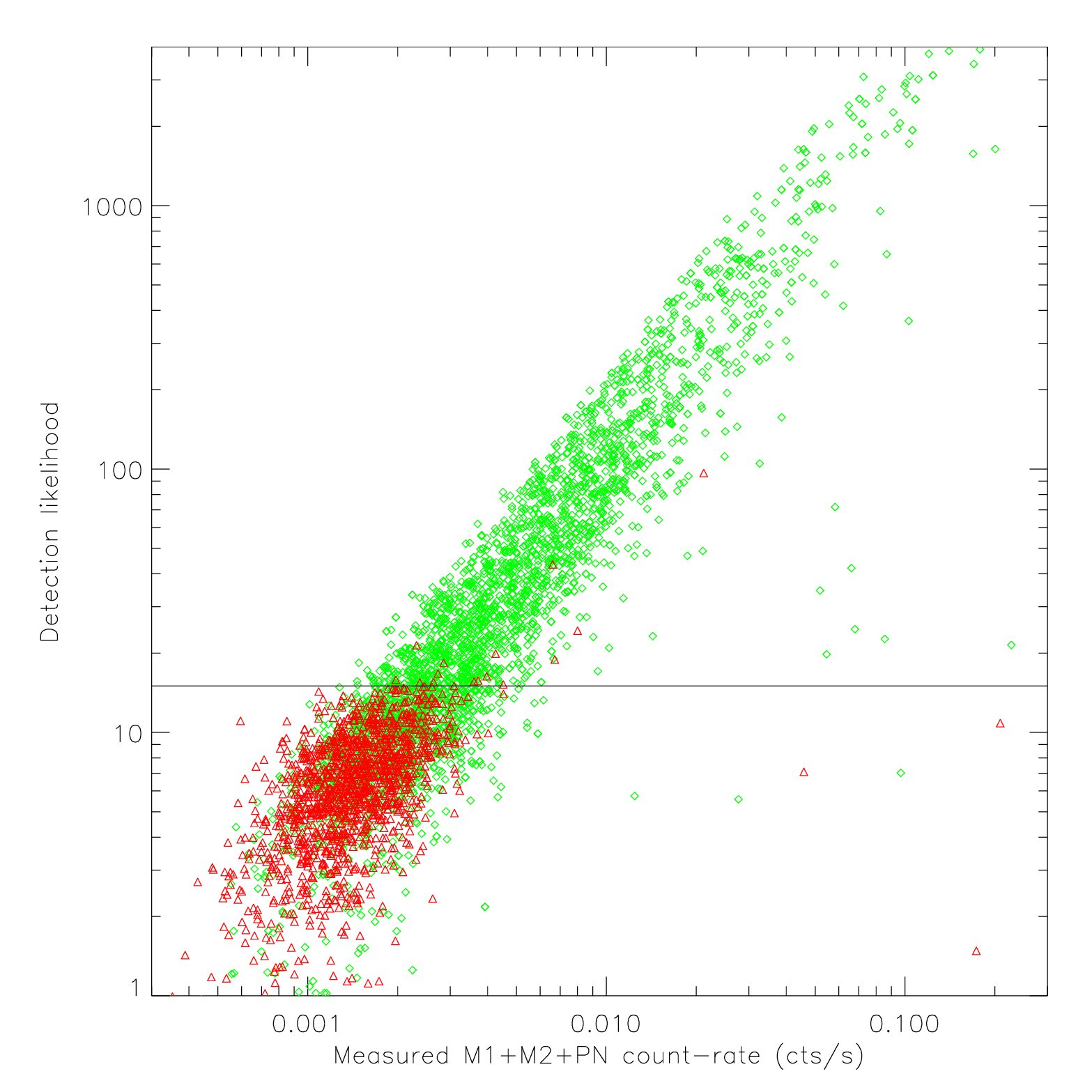}
  \hspace{.7cm}\includegraphics[width=7cm]{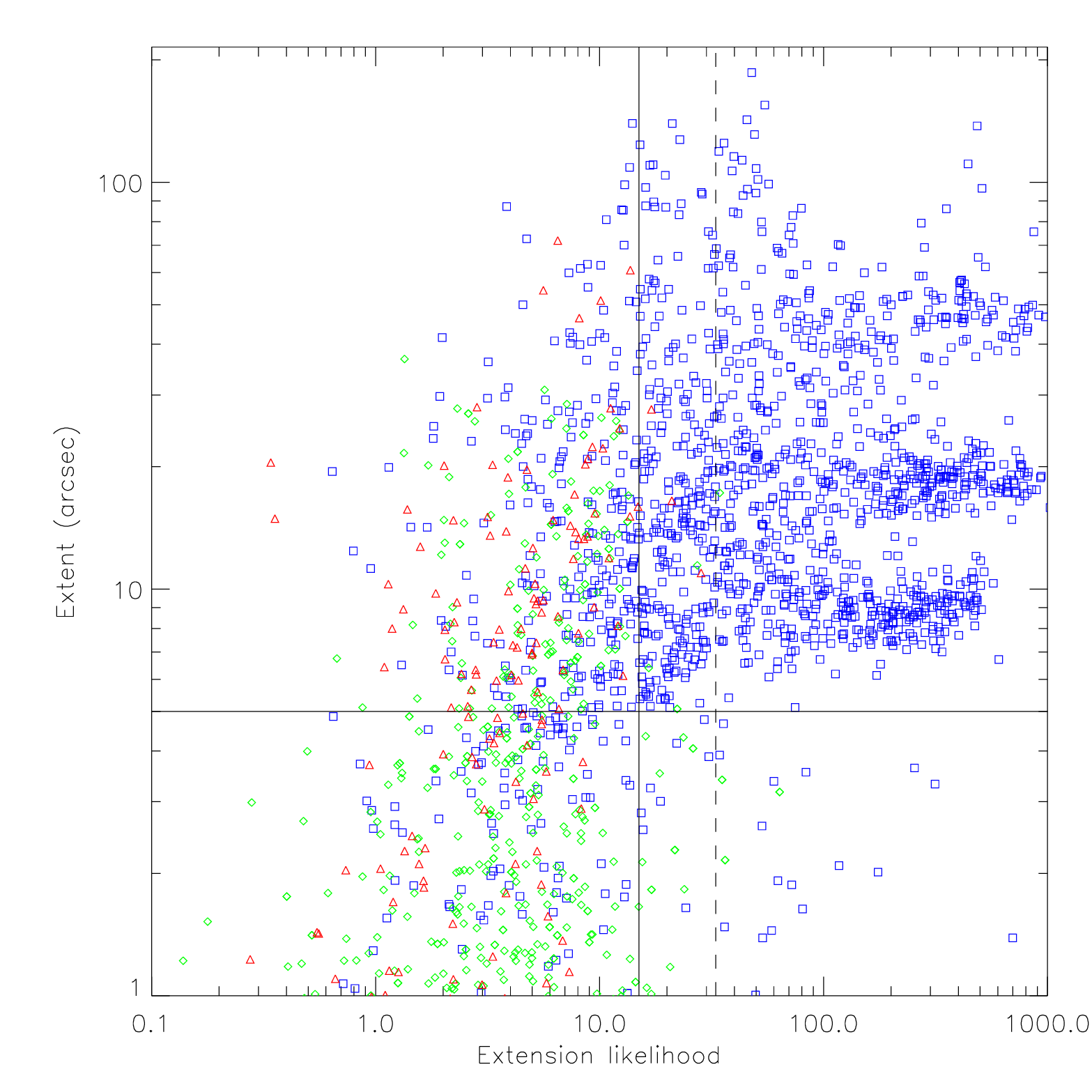}
  \rput(-13.2,6.1){\large (a)}
  \rput(-5.4,6.1){\large (b)}
}
\end{pspicture}
\caption{
Determination of the XMM-LSS pipeline selection 
criteria\label{class}. AGNs are displayed as green diamonds, galaxy 
clusters as blue squares. Red triangles stand for spurious detections.
Panel (a): Selection of point sources in the Count Rate - Detection 
Likelihood plane; the solid line at Likelihood=15 defines the point source sample.
Panel (b): cluster selection in the Extent - Extension Likelihood plane; 
the solid lines at Extent=5\arcsec\hspace{0.4mm} and Likelihood=15 define 
the C2 sample; the dashed line shows the extension 
likelihood criteria of the C1 sample.
}
\end{center}
\end{figure}

As a first step, we scanned the detection/extension statistic-extension 
space for the largest uncontaminated extended source sample. 
This is obtained for EXT $>$ 5\arcsec, EXT\_STAT $>$ 33, and extended 
fit DET\_STAT $>$ 32 simultaneously\footnote{ Note that from the 
definition of EXT\_STAT (see section \ref{stat} and Table \ref{xaminpar}), 
it is very unlikely for DET\_STAT to be lower than 32 if EXT\_STAT is 
greater than 33, except in the few rare cases where the point source 
fit crashed} (see Table \ref{xaminpar} for the definition of these 
parameters).  
From now on, we will refer to this sample as class 1 (C1) 
extended sources.
Figure \ref{class}a illustrates the main C1 selection process in the 
extension $-$ extension statistic plane.  .
\begin{table}
  \begin{center}
  \caption{Source selection criteria with the XMM-LSS pipeline\label{classtab}}
  \begin{tabular}{ll}
  \hline
  Classification \hspace{1.cm} & \hspace{1.3cm}Criteria\hspace{.7cm} \\
  \hline
  Class 1 extended &	Detection statistic$>$32,\\
  		   &	Extension statistic$>$33,\\
  	 	   &	Extension$>$5\arcsec \vspace{0.2cm}\\

  Class 2 extended &	Extension statistic$>$15,\\
 		   &    Extension$>$5\arcsec \vspace{0.2cm}\\

  Point source     &    Neither C1 nor C2,\\
  		   &    Detection statistic$>$15\\
  \hline
  \end{tabular}
  \end{center}
\end{table}

Due to our non-contamination requirement, the C1 sample naturally
excludes a number of extended sources (generally very low surface
brightness or more compact sources).  A less conservative sample
(required by the XMM-LSS cluster search in order to detect as many
valid sources as possible) can be obtained by relaxing the previous
criteria to EXT $>$ 5\arcsec, EXT\_STAT $>$ 15 and no DET\_STAT
constraint (see figure \ref{class}).  From the number of detections
matching these criteria in our point source simulations, we can
estimate that this class 2 sample (C2) contains less than one spurious
detection or misclassified point source every three pointings.

\begin{figure}
\begin{center}
\begin{pspicture}(0,0)(16,6.15)
\vbox{
  \hspace{.8cm}\includegraphics[width=7.5cm]{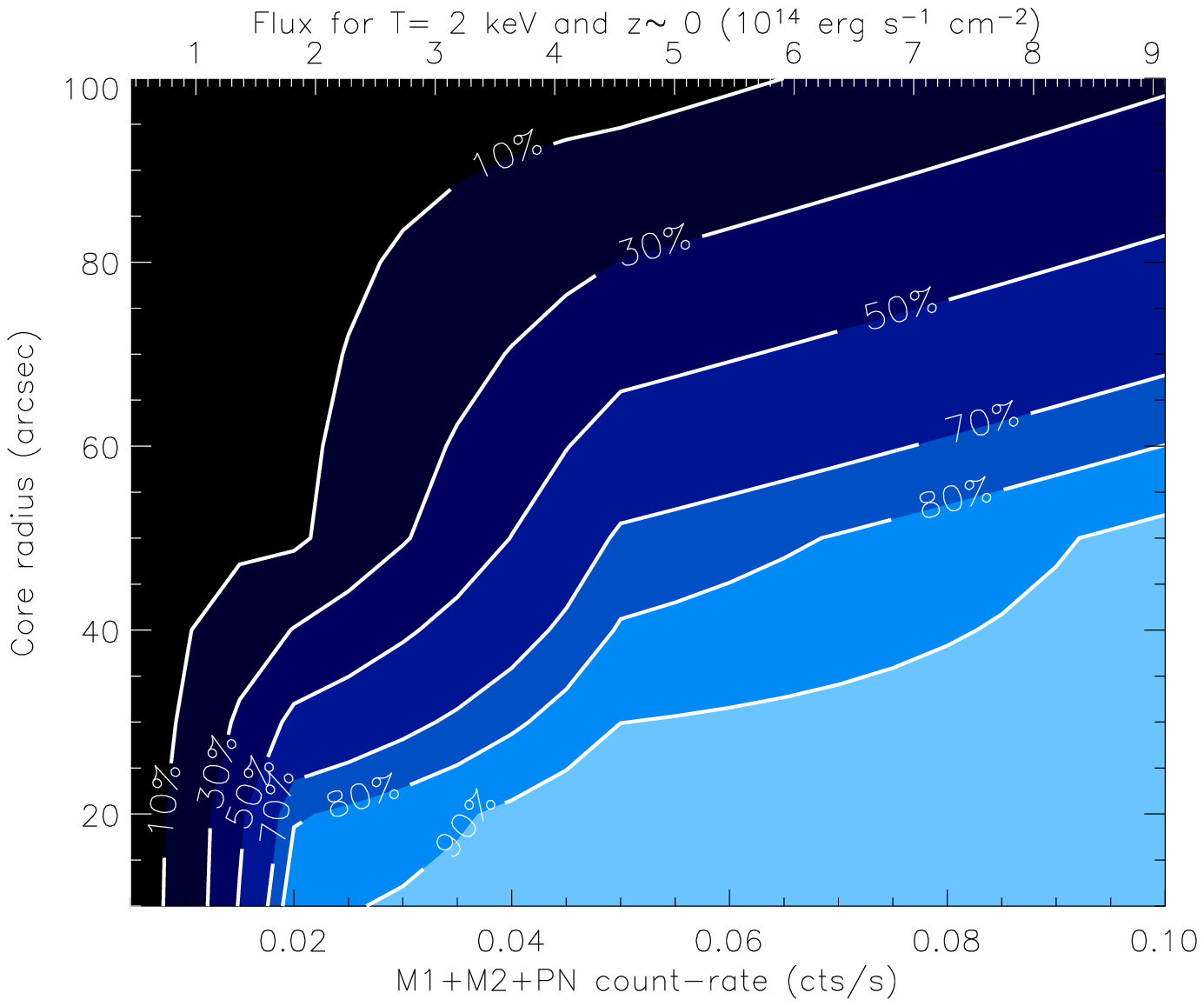}
  \includegraphics[width=7.5cm]{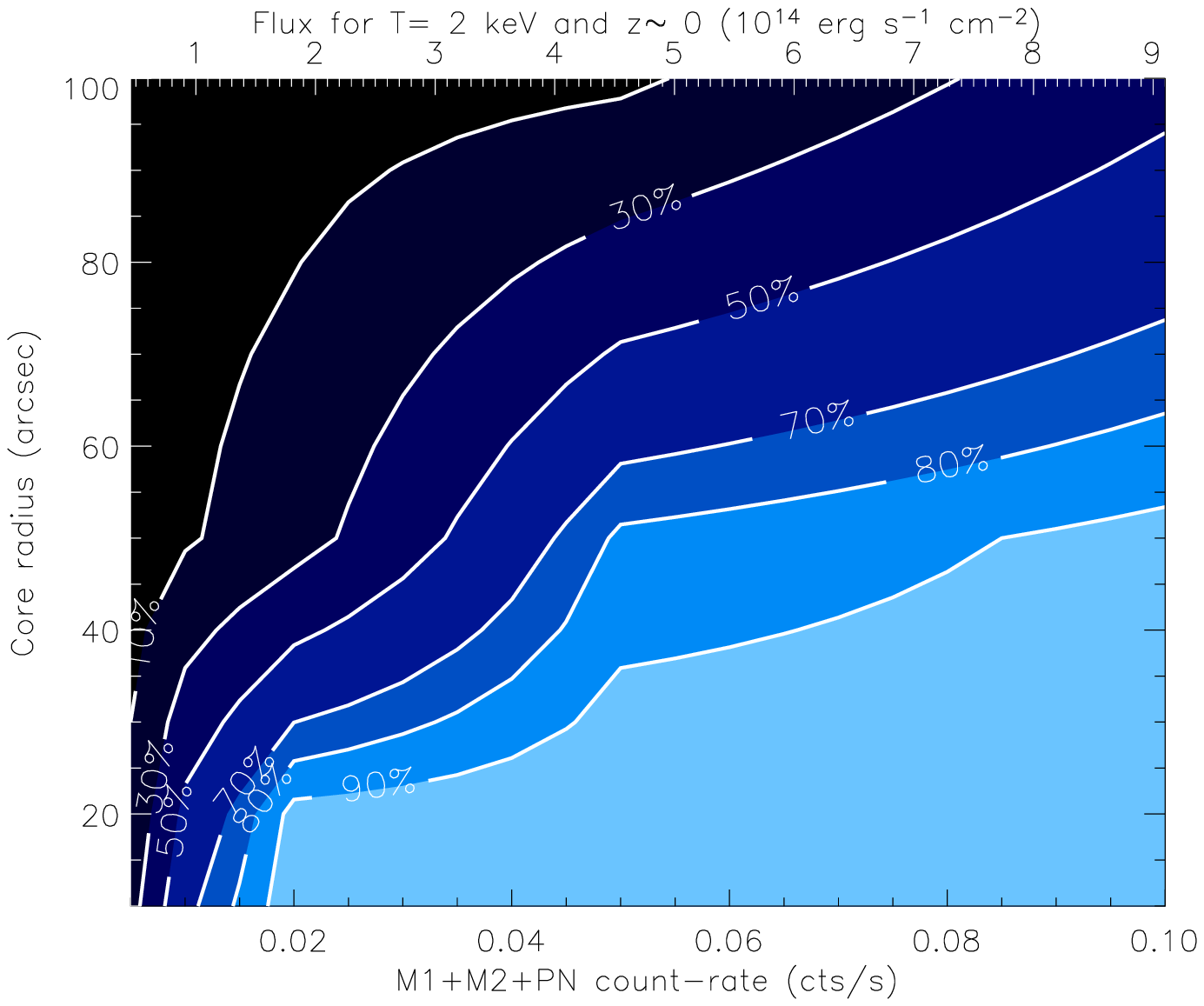}
  \rput(-13.8,4.8){\textcolor{white}{\large (a)}}
  \rput(-6.2,4.8){\textcolor{white}{\large (b)}}
}
\end{pspicture}
\caption{
Extended source detection efficiency of the XMM-LSS pipeline in $10^4$\,s exposures
as a function of source counts and core radius inside 10\arcmin\hspace{0.4mm} 
of the FOV.
\label{extdet} Panel (a): C1 sample. Panel (b): C2 sample.}
\end{center}
\end{figure}

The mean detection probabilities of extended sources within 10\arcmin\
of the FOV are presented for both C1 and C2 samples in figure
\ref{extdet}, as a function of count rate and apparent core radius.
As expected, this probability is higher within the C2 sample for low
surface brightness and faint compact sources.

\begin{figure}
\begin{center}
 \vspace{-.5cm}\includegraphics[width=6.5cm]{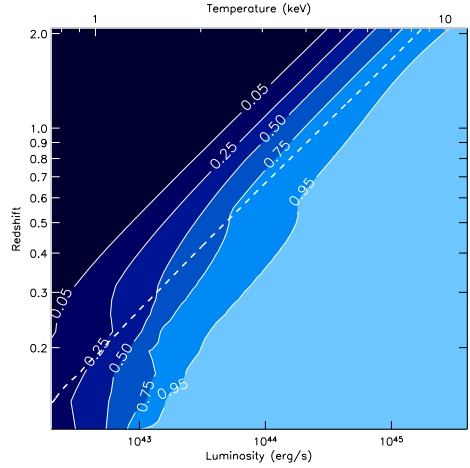}
\caption{
Probability of detecting a cluster located inside the central 10\arcmin\hspace{0.4mm}
of the FOV as a C2 source as a function of its redshift and luminosity given 
our cosmological model within $10^4$s pointings. An indicative flux limit of
$2\times10^{-14}$~erg.s$^{-1}$.cm$^{-2}$ is shown by the thick dashed line.\label{lzprob}}
\end{center}
\end{figure}

Note that detection efficiency is not a simple function of source flux
as is sometimes assumed in X-ray cluster surveys (see e.g. \citealt{rosati}),
but it varies significantly when considering different source sizes, and this 
should be modelled to interpret correctly the results of the XMM-LSS.
This impact of source extent on our detection capacity 
is illustrated by figure \ref{lzprob}, where the detection probability as a 
function of luminosity and redshift is shown for the C2 sample, assuming a 
canonical core radius of  $180h_{70}^{-1}$~kpc. 
At high redshift, where the angular distance is almost constant, our 
selection process closely resembles a flux limit, while the sensitivity 
drops at lower redshift. In this model, we find that roughly 90\% of the 
sources down to 3~keV are detected in C2 at $z$=0.5. This number falls 
to 50\% at a redshift of \hbox{0.9-1}. A cluster similar to Coma ($\sim$8~keV)
would always be detected at least as C2 up to a redshift of 1, and have more 
than 75\% probability of being detected at $z$=2.

\begin{table}
\begin{center}
 \caption{Contamination statistics predicted from the simulations for 
 each XMM-LSS pipeline source sample\label{cont}}
 \begin{tabular}{ccc}
 \hline
 Real source type &	Classification 	& $N_{Src}$/pointing\\
 \hline
 Point-like 	  & 	class 1	  	&	    0.0\\
 Point-like 	  & 	class 2	  	&	    0.17\\
 Spurious 	  & 	class 1	  	&	    0.0\\
 Spurious 	  & 	class 2	  	&	    0.10\\
 Spurious 	  & 	point-like    	&	    0.53\\
 \hline
 \end{tabular}
\end{center}
\end{table}

\subsection{Validation on real data}  
To further validate our selection criteria, we processed all available
XMM-LSS pointings and compared the pipeline output with our simulation
results. Our X-ray data currently consist of 51 XMM-NEWTON pointings;
19 of them (G pointings) were obtained from guaranteed-time
observation as part of a joint Li\`ege/Milan/Saclay program (XMDS,
\citealt{xmds}) and have $2\times10^4$\,s exposure; the remaining 32 are
$10^4$\,s long and were obtained with guest-observer time. Among these, 
three pointings (one G and two B) are unusable due to very high background
levels (probable solar flare contamination).

\subsubsection{Point sources}
In our 30 $10^4$\,s exposure pointings, we obtain on average 45.8 sources
per pointing with DET\_STAT $>$ 15 for the point source fit.  As a
comparison, taking into account the detection probabilities of figure
\ref{pntdet} and integrating over the $\log (N)-\log (S)$ of
\citet{moret} between $5\times10^{-16}$ and
$1\times10^{-11}$~erg~s$^{-1}$~cm$^{-2}$ yields on average 47.4
sources per pointing.  Though the matching is already satisfactory,
the remaining difference mostly reflects the lack of very bright
sources in the XMM-LSS area (probably due to cosmic variance)
identified by \cite{lssacf}.

We additionally cross-identified our sources with those of the
XMDS/VVDS $4\sigma$ catalogue \citep{xmds} which results from an
alternate analysis of the G fields (that mainly uses the standard
XMM-SAS procedures and is thus suitable only for point sources).  We
found a very good agreement with both detected sources and their
characteristics. {\it \texttt{Xamin}} count rates are always within
the error bars of XMM-SAS \texttt{emldetect} measurements for sources
that do not fall on CCD gaps. Moreover, our detection statistic
values are tightly correlated to their detection probability estimates.

\subsubsection{Extended sources}

Until now, the XMM-LSS optical spectroscopy follow-up program (see \citealt{xmmlss}) 
enabled us to confirm about 60 cluster candidates.
This allowed cross-checking the definition of our selection criteria obtained from the 
simulations against real data.

As regards the C1 sample, only genuine extended X-ray sources are detected, 
as expected, with no additional contamination. The majority ($\approx$85\%) 
of the extended C1 sources are clusters, the remainder being nearby galaxies.
A little contamination, around 0.5 false detections per pointing, is
observed in the C2 sample (in the present datasets, this amounts to
about 50\% of the newly-detected sources, once the nearby galaxies and
C1 clusters have been excluded). This false detection rate, while
still acceptable for a survey with optical follow-up, is slightly
higher than our estimates from simulations, and this is probably the
result of neglecting the AGN correlation function (thus lowering the
number of non-deblended close pairs of AGNs). Another possibility is
that we are detecting some AGNs that are included in cosmological
filaments with weak X-ray emission, which was not accounted for in the
simulations.

\subsubsection{Example runs on z$>$1 clusters}

We ran the pipeline on the archival XMM-NEWTON observation 
0111790101, for which detection of the highest redshift X-ray 
cluster to date, XMMUJ2235.3-2557, at $z\sim1.4$ was reported 
\citep{mullis1p4}.
The observation was performed in the MEDIUM FILTER and 
EPIC-PN small window mode so that the source, located 7.7\arcmin\ from 
the optical axis, is only observed in the EPIC-MOS detectors.
Using the full $4.5\times10^4$\,s exposure of the pointing ($\approx$ 
$3\times10^4$\,s at the 
source position), the cluster is easily identified as a C1 extended 
source.
We further simulated the XMM-LSS observing conditions by analyzing only the 
first $10^4$\,s of the observation. 
XMMUJ2235.3-2557 is still detected with extended fit DET\_STAT=93.8, 
EXT\_STAT=31.1 and EXT=9.8, as a C2 extended source, at the limit of the C1 
parameter space, and would therefore have been detected as C1 in the exact 
XMM-LSS observing conditions (i.e. using THIN FILTER and with EPIC-PN data 
available). 
The ease with which this high redshift cluster is detected is mainly due
to its apparent brightness: $\sim$~220 (resp. 70) photons were 
available in the $4.5\times10^4$\,s ($10^4$\,s) exposure. 
For comparison, we note that the z=1.22 cluster XLSSJ022302.6-043621
detected by \citet{bremer} in the XMM-LSS survey is classified as a C2 source
(EXT=5.4, EXT\_STAT=15.4, and DET\_STAT=51.4) with only 58 photons available 
for the fit.  
  
\section{The XMM-LSS selection function}
\label{cosmo}
Our simulation programs provide us with tools to compute 
the XMM-LSS selection function. We can derive the detection probability 
as a function of source characteristic for any exposure time, 
background level, and position on the detector.

Figure \ref{pntdet} shows the point-source detection probability
inside a radius of 10\arcmin\vspace{0.7mm} from the mean optical axis
as a function of flux. From this, a direct estimate of our mean sky
coverage can be obtained.
 
For a given cosmology, a galaxy cluster of given luminosity,
temperature, physical extent and redshift can be described by an
angular core radius and a detected XMM count rate, for which figure
\ref{extdet} gives the detection probability for C1 and C2. We are
therefore now able to properly describe our galaxy cluster selection
process.

As an illustration, we compute below the expected redshift 
distribution of C1 and C2 clusters in $\Lambda$CDM cosmology.    

\subsection{Cosmological model}

In the following, the cosmological parameters that determine the 
dynamics and content of the universe are set to WMAP values 
\citep{wmap} namely: \\
$H_0=71$~km~s$^{-1}$~Mpc$^{-1}$, $\Omega_m=0.27$, $\Omega_\Lambda=0.73$, 
$\Omega_b=0.044$, $n=0.93$ and $\sigma_8=0.84$.

\subsubsection{Mass function}

The shape of the linear power spectrum $P(k)$ is computed at z=0 
using the initial power law dependency in $k^n$ and the transfer 
function from \citet{BBKS}. The influence of baryons on the transfer
function was modelled using the shape parameter:
\begin{equation} 
  \Gamma=\Omega_m h\times exp\left[-\Omega_b\left(1 + \sqrt{2h}/\Omega_m\right)\right]
\end{equation} 
introduced by \citet{sugi}, and the overall spectrum is normalized to 
$\sigma_8$.

Then at each redshift value on a fine grid: $P(k)$ is evaluated from 
its z=0 value using the linear growth factor from \citet{carroll} 
and $\sigma(M)$ is deduced. The comoving halo number density as a 
function of mass, $dn/dm(z)$, is computed using the \citet{ShTormen}
mass function.

This common procedure to determine the halo mass function has been
largely tested on numerical simulations and is known to provide 
accurate predictions as long as one defines the mass of the haloes 
to be the one included inside $r_{200b}$, the radius that encloses an 
overdensity of 200 with respect to the mean background density.

\subsubsection{Applying the selection function}
Knowledge of the cluster scaling relations is needed to predict
the temperature and luminosity of these haloes and compute XMM 
count rates. 
Unfortunately, one generally doesn't have access to the mass in 
$r_{200b}$ from the X-ray data, and a halo profile model is required 
in order to convert the mass function to another mass definition.

For this purpose, we used NFW profiles with scaling radius $r_s$
provided by the model of \citet{bullock} which relates 
$r_s$ to the virial mass of the halo through the concentration 
parameter $c=r_{vir}/r_s$\footnote{Note that we also tested the model 
of \cite{eke} and found a change in the redshift distribution of our 
C1/C2 samples lower than 10\%}.
The conversion itself is performed using the formulae provided by 
the appendix of \citet{HuKravtsov}.

The emission-weighted gas temperature is derived using the local 
$M_{200}$-$T$ relation of \citet{arnaud2005}, i.e. a slope of 
$\alpha=1.49$, valid for clusters with $T>4$~keV. At lower temperatures,
we added a gradual steepening of the correlation ($\alpha=1.85$ below 
4~keV and $\alpha=2$ below 2~keV) as indicated by several recent works 
(see e.g. \citealt{FRB}). No evolution of the $M_{200}$-$T$ relation
with redshift was supposed. As an arbitrary condition to be considered 
as a group or cluster, we subsequently removed all haloes with $T<1$~keV.   

Bolometric luminosities are then computed using the $L_X$-$T$ relation of 
\citet{arnaudevrard} with no evolution. Though there is some evidence 
that the local $L_X$-$T$ relation also steepens at low $T$, this seems to be 
important only for $T\leq 1$~keV (Ponman et al., in preparation) and is 
consequently ignored.  

The total XMM-NEWTON EPIC count rate is estimated using an APEC
\footnote{http://cxc.harvard.edu/atomdb/} thermal plasma 
emission model \citep{APEC} with neutral hydrogen absorption as modelled 
by \citet{wabs} using fixed column density of $2.6\ 10^{20}$~cm$^{-2}$
(representative of our field) folded through the EPIC response 
matrices for the THIN filter in accordance with our observing mode. 

The selection function is finally applied assuming a constant physical 
core radius of $180h_{70}^{-1}$~kpc.

\subsection{Results}

\begin{figure}
\begin{center}
\begin{pspicture}(0,0)(16,5.6)
\vbox{
  \includegraphics[width=7.3cm]{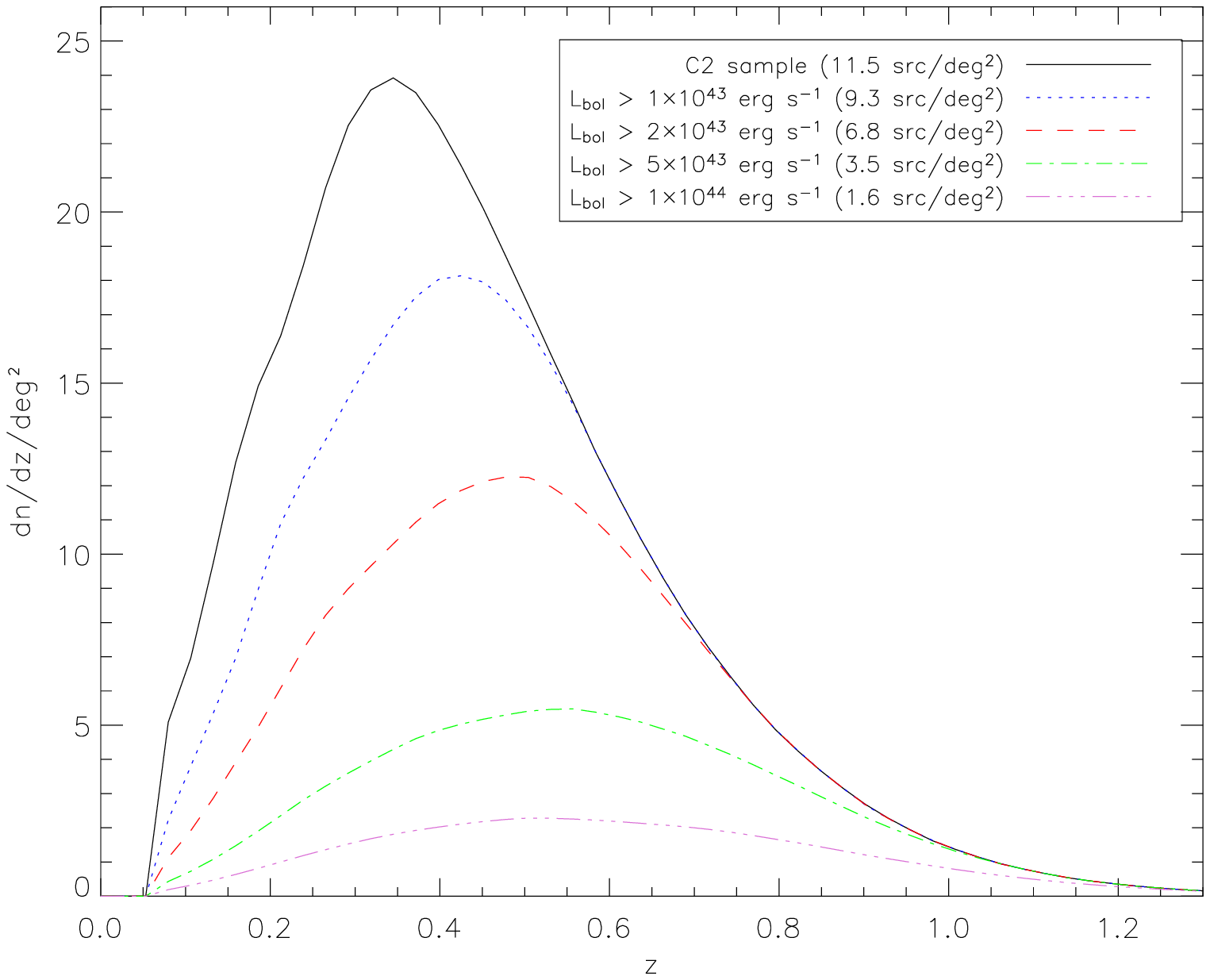}
  \includegraphics[width=7.3cm]{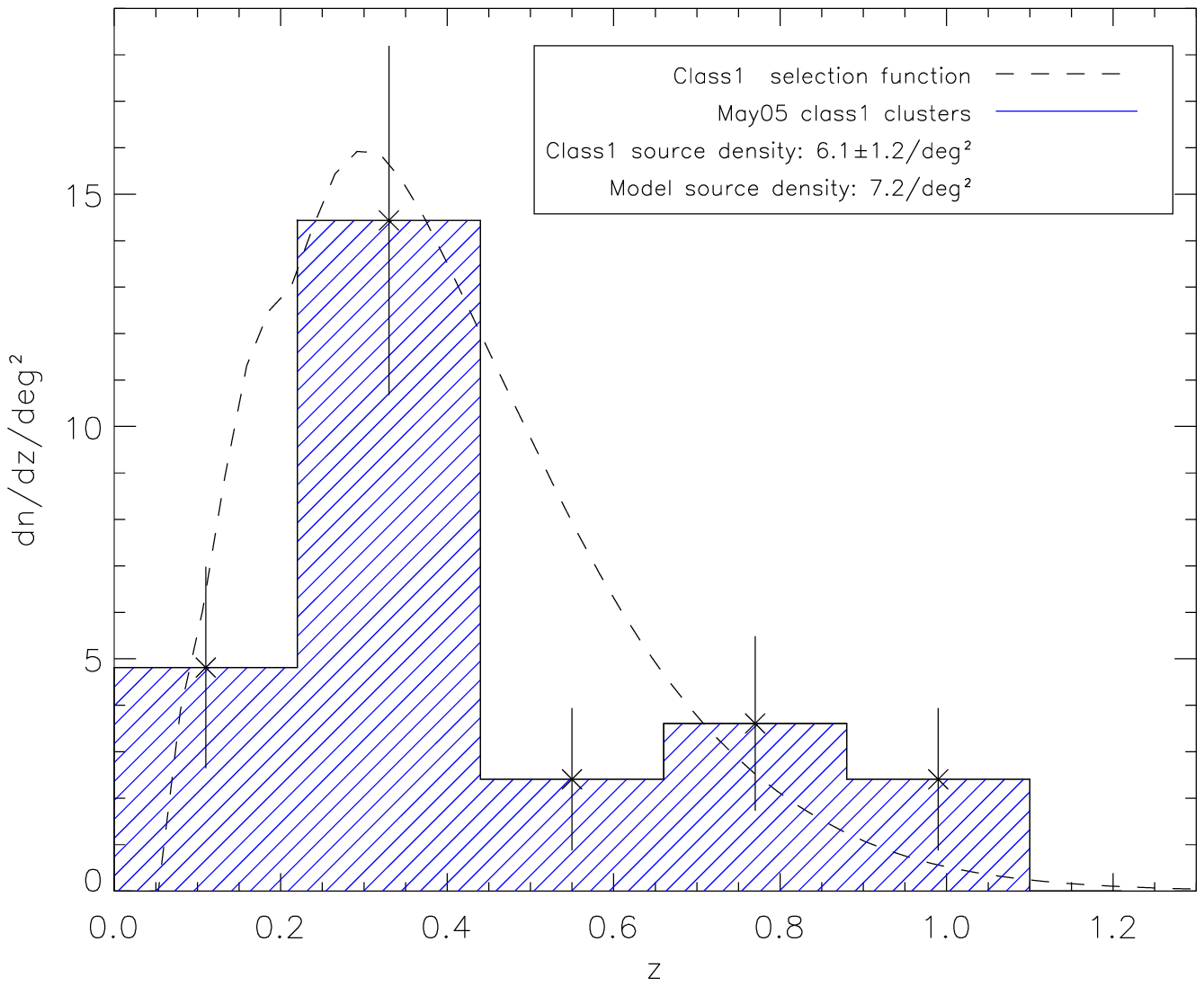}
  \rput(-13.2,4.85){\large (a)}
  \rput(-5.68,4.85){\large (b)}
}
\end{pspicture}
\caption{
Cosmological expectations of the C1 and C2 samples for sources with $T>$~1~keV. Panel (a): Luminosity and redshift distribution of the C2 sample. 
Panel (b): redshift distribution of the observed C1 sources (29 sources, 24 with redshifts) 
compared to the $\Lambda$CDM expectations. \label{dndz}}
\end{center}
\end{figure}

Using this simplified model and the selection functions obtained from the
simulations, we find that:
 \begin{itemize}
  \item The C2 sample should contain roughly 12 clusters per deg${^2}$.
  When the XMM-LSS is complete, it will thus constitute the deepest 
  X-ray selected galaxy cluster sample over a wide area. 
  \item The C1 sample should contain some 7 clusters per deg${^2}$.
  While this source density is a bit lower than for C2, this selection
  process can be applied to the whole of the XMM archive, regardless
  of expensive and time consuming optical spectroscopy follow-up, as
  the sample is effectively uncontaminated.
 \end{itemize}
The expected redshift distribution for both samples is shown in figure 
\ref{dndz}. Panel (a) of that same figure also gives an idea of the luminosity 
distribution of the C2 sample. 

To validate these results, we compared them with the redshift 
distribution of the observed C1 clusters. The sample contains 29 sources 
of which 24 have already been spectroscopically confirmed.
Assuming that the 5 missing sources ($\sim$~17\% of the sample) will not alter 
significantly the current distribution, we find very good overall agreement 
with our prediction (fig. \ref{dndz}b).

\begin{figure}
\begin{center}
\begin{pspicture}(0,0)(16,5.7)
\vbox{
  \includegraphics[width=7.3cm]{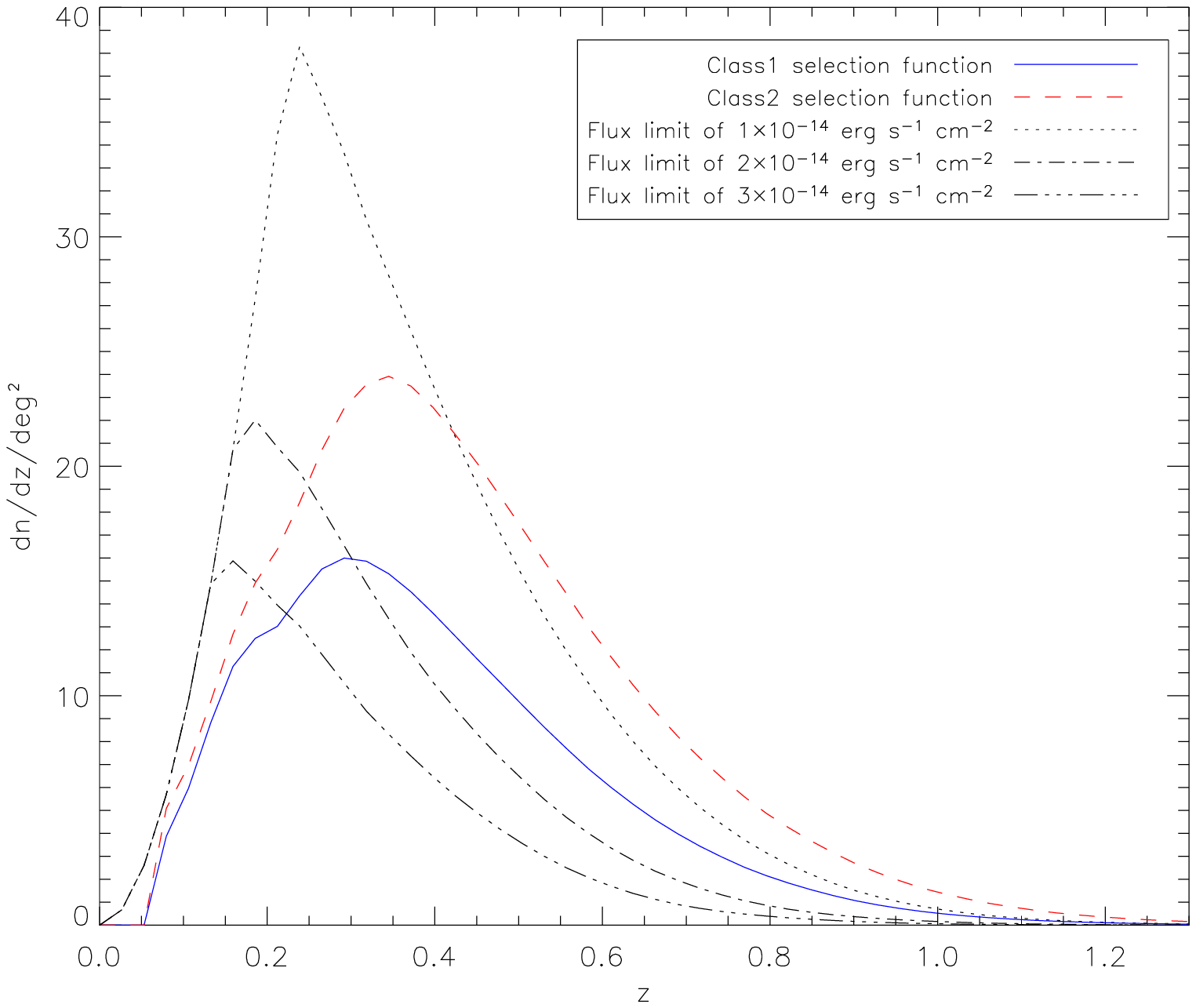}
  \includegraphics[width=7.3cm]{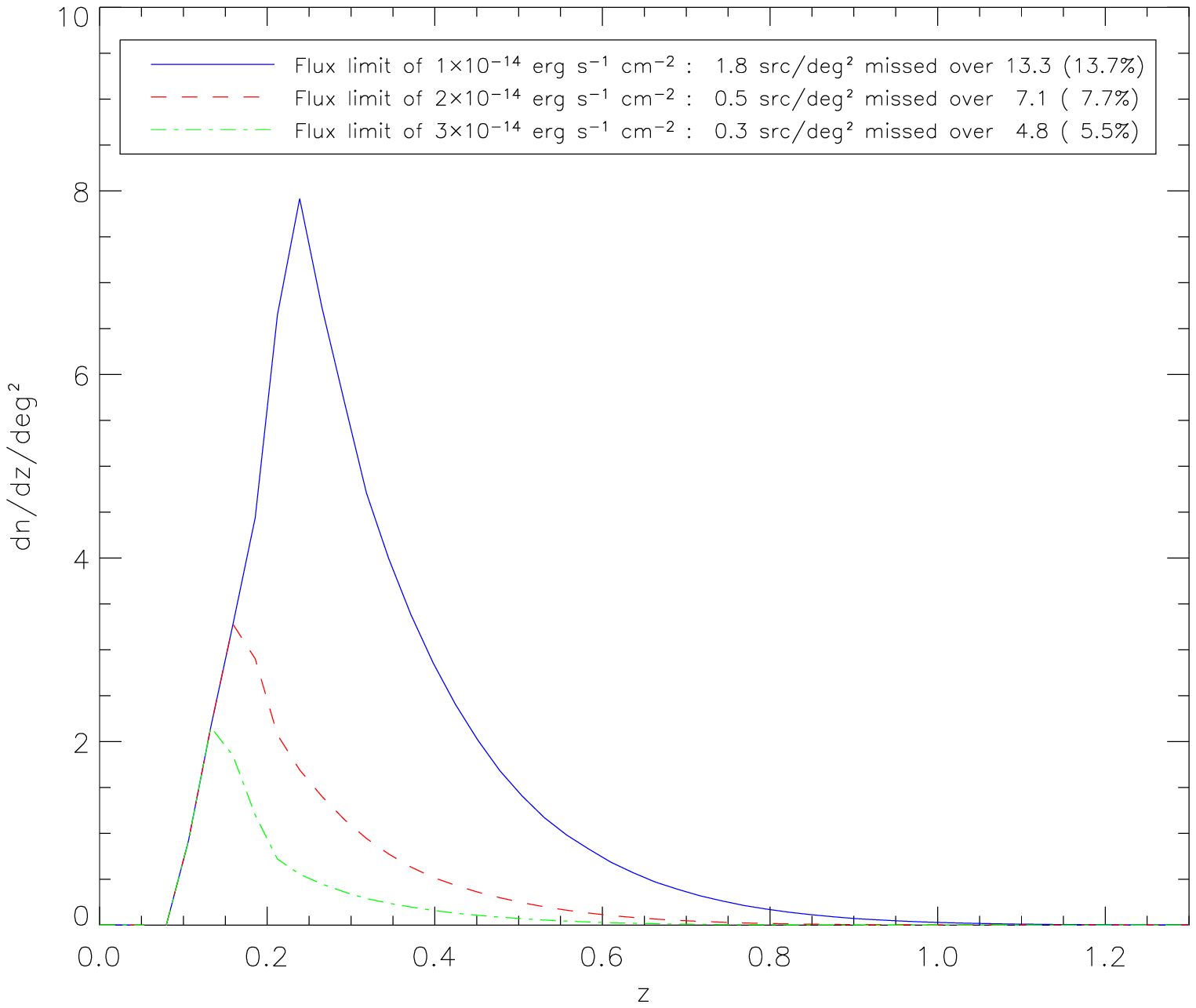}
  \rput(-8.12,1.4){\large (a)}
  \rput(-.7,1.4){\large (b)}}
\end{pspicture}

\caption{\label{flmissed}
Comparison of our source selection process with the common assumption of a 
flux limited sample for sources with $T>$~1~keV. 
Panel (a): expected $dn/dz$ for class 1 (blue) and class 2 (red) compared to 
flux limited surveys; on average, we detect higher redshift clusters 
than flux limited surveys with the same source density. 
Panel (b): redshift distribution of the sources not detected by the 
first pass  (\texttt{MR/1} + \texttt{SExtractor}) above several flux 
limits, assuming the source population generated by our simple cosmological
model (we miss the low end  of the luminosity function).
}
\end{center}
\end{figure}

A further interesting result, already outlined in section
\ref{extclassdef} and fig. \ref{extdet}, is that our selection process
doesn't reproduce a flux limited sample, especially at $z<0.6$ where the 
change in angular distance is significant (see figure \ref{flmissed}a). 

This point is further illustrated in fig. \ref{flmissed}b where we 
investigate our detection efficiency as a function of cluster flux. 
This shows, for the assumed cluster population, the a priori 
impossibility of constructing a flux limited sample from our primary 
catalogues, even accepting a substantial contamination level, unless a 
very high flux limit is set.
In the present study, \texttt{SExtractor} is run on optimally filtered
images (retaining only significant structures above $3\sigma$) with a 
very sensitive detection threshold and our results suggest that we have 
reached the limit of the data. This therefore challenges any further 
attempt aiming at defining deep flux limited samples with XMM.

\subsection{Limitations of the present model} 

Although the matching between this simple model and our data (as shown in
fig. \ref{dndz}b) is impressive, one should keep in mind 
that some ingredients of the model are still uncertain (and this 
is precisely the purpose of the XMM-LSS to try to constrain them).

In particular, while the evolution of the $M_{200}$-$T$ relation is still
unknown, there seems to be indication of a positive evolution as
predicted by self-similar models (see e.g. \citealt{ettoriscaling},
\citealt{benscaling}). However none of these studies is probing our
range of temperature and redshift, and the influence of
non-gravitational processes can well alter this behaviour in the group
regime, thus the use of the simplest non-evolving relations.

Also, in order to properly take into account the varying gas distribution 
with cluster mass, our assumption of a fixed core radius may seem too
simple and one would have to consider lower $\beta$ values for the groups 
as indicated by observations (e.g. \citealt{GEMS}).
However such data are generally largely dominated by scatter and there is 
currently no well-established scaling relation for these global trends.

Finally, a large fraction of the observed scatter on all these scaling
relations is intrinsic to the source properties and results from the
complex process of hierarchical merging in cold dark matter
cosmologies and feedback from non-gravitational activity. 

These are a number of caveats that neeed to be taken into account in the 
interpretation of such a small sample of low temperature systems. 
In a forthcoming paper \citep{lsscosmo}, where we will present the full 
cluster catalogue, we shall further discuss the effect of the various 
cluster scaling laws and evolution schemes on the $dn/dz$ using as input 
our $L-T$ relation for groups at redshift around 0.5.

\section{Summary and conclusions}

We have described the procedure that we developed to analyse the
$1\times10^4-2\times10^4$\,s XMM images of the XMM-LSS survey. The main motivation of
this work is the need for assembling a sample of clusters of
galaxies out to a redshift of unity with controlled selection
effects, suitable for cosmological and evolutionary studies.
The resulting pipeline consequently combines multi-resolution 
wavelet filtering (\texttt{MR1}) to reach the source detection 
limit, with a subsequent maximum likelihood analysis (\texttt{Xamin}) 
to characterize the source properties.

The performances of the adopted procedure have been duly
tested by means of extensive image simulations: either reproducing
all instrumental and astrophysical effects, or injecting extended
and point-like sources into already existing pointings. This allowed
us to investigate the ultimate capabilities such as: resolving
power, cluster detectability and characterization as a function of
flux and apparent size, photometric accuracy. In this respect, our
package constitutes a significant improvement over the standard SAS
and the XMDS procedure \citep{xmds}, specially for the extended
source analysis.

Moreover, the \texttt{Xamin} output parameter space, densely scanned
by the simulations, provides a powerful means to interpret the
detected sources. In this way, we are able to define two classes of
extended sources: the C1 class which is basically uncontaminated by
misclassified point-like sources, and the C2 class allowing for some
50\% contamination. This selection process, derived from the
simulations, has been subsequently checked and validated against the
current XMM-LSS sample of spectroscopically confirmed galaxy clusters.

Finally, considering a canonical power spectrum combined with a simple
halo model providing $n(M,z)$ and simple cluster scaling laws ($M$-$T$-$L$) 
in a $\Lambda$CDM cosmology allowed us to predict the $dn/dz$ distribution
of the C1 cluster population. Comparison with our current C1 data
sample shows a very good agreement. From this, we infer that our goal
of producing a cluster sample with controlled selection effects is
fulfilled at this stage. An important point to be further emphasized
is that the resulting sample is not flux limited - a concept that is
anyway not rigorously applicable when dealing with extended sources 
spanning a wide range in flux and size.

The way the C1 class is defined allows us to construct a purely X-ray
selected cluster sample with a high number density of $\sim 7$/deg$^2$
in the redshift range [0-1.2]. 
Moreover, an unprecedented density of $\sim 12$/deg$^2$ can be obtained 
with the C2 sample which includes objects down to a flux of 
$\sim~5\times10^{-15}$~erg~s$^{-1}$~cm$^{-2}$.  
This opens the door to the routine construction of unbiased cluster 
samples from XMM images.

In the very near future, with the compilation of the full XMM-LSS
cluster sample over the currently existing 5 deg$^2$, we shall refine
the cosmological modelling of the observed $dn/dz$ \citep{lsscosmo}.  In
particular, we shall further investigate the effect of varied
evolution schemes of the scaling relation, and assumptions on cluster
sizes and shapes (including scatter on these average trends).  Both
aspects are especially relevant for the $T<$~2~keV groups out to $z\sim$~0.5,
a population that the XMM-LSS is for the first time unveiling and that
constitute the bulk of our sample. Noting that the C1 cluster sample
is almost identical to the sample for which we can measure a
temperature \citep{D1paper}, we shall also be in a position to
constrain the evolution of the $L_X$-$T$ relation.

The combined $dn/dz$, $L_X$-$T$, and shape-modelling will provide very useful
constrains on numerical simulations, the missing link between the
theoretical parameter $M$ and the observable $L_X$, and consequently a
self-consistent description of the building blocks of the present day
clusters.

\section*{acknowledgements}
This paper was based on observations obtained with XMM-NEWTON, an 
ESA science mission with instruments and contributions directly 
funded by ESA Member States and NASA. 
The simulations were performed at the CNRS `Centre de Calcul de l'IN2P3'
located in Lyon, France. 
The authors would like to thank Jean Ballet for sharing with us 
his deep knowledge of statistics and XMM-NEWTON calibrations,
and Pierrick Micout for his help regarding the use of the CC-IN2P3. 
We are also grateful to Jean-Paul Le F\`evre for developing and
monitoring our cluster database (the L3SDB\footnote{http://l3sdb.in2p3.fr:8080/}) 
thus helping us to handle our large data sets, and to Monique Arnaud for 
useful discussions and advices regarding local X-ray cluster scaling relations.

\bsp 

\onecolumn

\begin{thebibliography}{}
\bibitem[\protect\citeauthoryear{Arnaud \& Evrard}{1999}]{arnaudevrard} Arnaud, M., \&
Evrard, A.~E.\ 1999, MNRAS, 305, 631
\bibitem[\protect\citeauthoryear{Arnaud et al.}{Arnaud, Pointecouteau \& Pratt}{2005}]{arnaud2005} 
Arnaud, M., Pointecouteau, E., \& Pratt, G.~W.\ 2005, A\&A, 441, 893
\bibitem[\protect\citeauthoryear{Bardeen et al.}{1986}]{BBKS} Bardeen, J.~M., Bond,
J.~R., Kaiser, N., \& Szalay, A.~S.\ 1986, ApJ, 304, 15
\bibitem[\protect\citeauthoryear{Bertin \& Arnouts}{1996}]{sextra}
Bertin, E., \& Arnouts, S. 1996, A\&AS, 117, 393 ({\bf \texttt{SExtractor}})
\bibitem[\protect\citeauthoryear{Bremer et al.}{2006}]{bremer}
Bremer, M.~N., Valtchanov, I., Willis, J., et al. (2006), submitted to MNRAS
\bibitem[\protect\citeauthoryear{Bullock et al.}{2001}]{bullock} Bullock, J.~S., Kolatt,
T.~S., Sigad, Y., Somerville, R.~S., Kravtsov, A.~V., Klypin, A.~A.,
Primack, J.~R., \& Dekel, A.\ 2001, MNRAS, 321, 559
\bibitem[\protect\citeauthoryear{Cash}{1979}]{cash} 
Cash, W.\ 1979, ApJ, 228, 939
\bibitem[\protect\citeauthoryear{Carroll et al.}{1992}]{carroll} Carroll, S.~M., Press,
W.~H., \& Turner, E.~L.\ 1992, ARA\&A, 30, 499
\bibitem[\protect\citeauthoryear{Cavaliere \& Fusco-Femiano}{1976}]{cavaliere1976} Cavaliere,
A., \& Fusco-Femiano, R.\ 1976, A\&A, 49, 137
\bibitem[\protect\citeauthoryear{Chiappetti et al.}{2005}]{xmds} Chiappetti, L., Tajer, M., 
Trinchieri, G.\ et\ al.\ 2005, A\&A, 439, 413
\bibitem[\protect\citeauthoryear{Eke, Navarro \& Steinmetz}{2001}]{eke} 
Eke, V.~R., Navarro, J.~F., \& Steinmetz, M.\ 2001, ApJ, 554, 114
\bibitem[\protect\citeauthoryear{Ettori et al.}{2004}]{ettoriscaling} Ettori, S., Tozzi, P.,
Borgani, S., \& Rosati, P.\ 2004, A\&A, 417, 13
\bibitem[\protect\citeauthoryear{Finoguenov, Reiprich \& B\"ohringer}{Finoguenov et al.}{2001}]{FRB} Finoguenov, A.,
Reiprich, T.~H.,\ \&\ B\"ohringer, H.\ 2001, A\&A, 368, 749
\bibitem[\protect\citeauthoryear{Gandhi et al.}{2006}]{lssacf}Gandhi, P., Garcet, O., Disseau, L.\ et\ 
al.\ 2006, A\&A in press
\bibitem[\protect\citeauthoryear{Hu \& Kravtsov}{2003}]{HuKravtsov} Hu, W., \& Kravtsov,
A.~V.\ 2003, ApJ, 584, 702
\bibitem[\protect\citeauthoryear{Infante}{1987}]{SEinfante}
Infante, L. 1987, A\&A, 183, 177
\bibitem[\protect\citeauthoryear{Kron}{1980}]{SEkron}
Kron, R.G. 1980, ApJS, 43, 305
\bibitem[\protect\citeauthoryear{Maughan et al.}{2005}]{benscaling}
Maughan, B.~J., Jones, L.~R., Ebeling, H., \& Scharf, C.\ 2006, MNRAS, 365, 509 
\bibitem[\protect\citeauthoryear{Moretti et al.}{2003}]{moret} 
Moretti, A., Campana, S., Lazzati, D., \& Tagliaferri, G.\ 2003, ApJ, 588, 696 
\bibitem[\protect\citeauthoryear{Morrison \& McCammon}{1983}]{wabs} 
Morrison, R., \& McCammon, D.\ 1983, ApJ, 270, 119
\bibitem[\protect\citeauthoryear{Mullis et al.}{2005}]{mullis1p4} Mullis, C.~R., Rosati,
P., Lamer, G., B{\"o}hringer, H., Schwope, A., Schuecker, P., \&
Fassbender, R.\ 2005, ApJL, 623, L85
\bibitem[\protect\citeauthoryear{Osmond \& Ponman}{2004}]{GEMS} Osmond, J.~P.~F., \&
Ponman, T.~J.\ 2004, MNRAS, 350, 1511
\bibitem[\protect\citeauthoryear{Pierre et al.}{2004}]{xmmlss}
Pierre, M., Valtchanov, I., Altieri, B.\ et\ al.\ 2004,
JCAP, 9, 11
\bibitem[\protect\citeauthoryear{Pierre et al.}{2005}]{D1paper}
Pierre, M., Pacaud, F., Duc, P.-A. et al. 2006, MNRAS in press
\bibitem[\protect\citeauthoryear{Pacaud et al.}{2006}]{lsscosmo}
Pacaud, F.\ et\ al.\ 2006, in preparation
\bibitem[\protect\citeauthoryear{Pratt \& Arnaud}{2002}]{pratt2002}
Pratt, G., \& Arnaud, M. 2002, A\&A, 394, 375
\bibitem[\protect\citeauthoryear{Press et al.}{1992}]{NumRec} Press, W.~H., Teukolsky, S.~A.,
Vetterling, W.~T., \& Flannery, B.~P.\ 1992, Cambridge: University Press, 2nd ed. 
\bibitem[\protect\citeauthoryear{Read \& Ponman}{2003}]{read2003} 
Read, A.~M.,~\& Ponman, T.~J.\ 2003, A\&A, 409, 395
\bibitem[\protect\citeauthoryear{Rosati et al.}{1998}]{rosati} Rosati, P., della Ceca,
R., Norman, C., \& Giacconi, R.\ 1998, ApJL, 492, L21
\bibitem[\protect\citeauthoryear{Spergel et al.}{2003}]{wmap} Spergel, D.~N.,
Verde, L., Peiris, H.~V. et al.\ 2003, ApJS, 148, 175
\bibitem[\protect\citeauthoryear{Sheth \& Tormen}{1999}]{ShTormen} Sheth, R.~K., \&
Tormen, G.\ 1999, MNRAS, 308, 119
\bibitem[\protect\citeauthoryear{Smith et al}{2001}]{APEC} Smith, R.~K., Brickhouse,
N.~S., Liedahl, D.~A., \& Raymond, J.~C.\ 2001, ApJL, 556, L91 ({\bf \texttt{APEC}})
\bibitem[\protect\citeauthoryear{Starck, Murtagh \& Bijaoui}{Starck et al.}{1998}]{mr1}
Starck, J.-L., Murtagh, F., \& Bijaoui, A. 1998, 
Image Processing and Data Analysis: 
The Multiscale Approach 
(Cambridge Univ. Press) (UK) ({\bf \texttt{MR/1}})
\bibitem[\protect\citeauthoryear{Starck \& Pierre,}{1998}]{starckpierre}
Starck, J.-L., \& Pierre, M. 1998, A\&AS, 128, 397
\bibitem[\protect\citeauthoryear{Sugiyama}{1995}]{sugi} 
Sugiyama, N.\ 1995, ApJS, 100, 281
\bibitem[\protect\citeauthoryear{Valtchanov, Pierre \& Gastaud}{Valtchanov et al.}{2001}]{val2001}
Valtchanov, I., Pierre, M., \& Gastaud, R. 2001, A\&A, 370, 689
\bibitem[\protect\citeauthoryear{Vikhlinin et al.}{1998}]{160sqdeg} 
Vikhlinin, A., McNamara, B.~R., Forman, W., Jones, C., Quintana, H., 
\& Hornstrup, A.\ 1998, ApJ, 502, 558
 \end{thebibliography}
\end{document}